\documentclass[twocolumn, times, tighten]{aastex62}
\usepackage{amsmath}

\shortauthors{Ben\'itez-Llambay, P., Krapp, L. \& Pessah, M. E.}

\begin{document}

\title{\sc{Asymptotically stable numerical method for multispecies momentum transfer: Gas and Multifluid dust test suite and implementation in FARGO3D}}

\shortauthors{Ben\'itez-Llambay, P., Krapp, L. \& Pessah, M. E.}
\shorttitle{Multispecies momentum transfer}
\author{Pablo Ben\'itez-Llambay}
\email{pbllambay@nbi.ku.dk} 
\author{Leonardo Krapp}
\email{krapp@nbi.ku.dk}
\author{Martin E. Pessah}
\affil{Niels Bohr International Academy, Niels Bohr Institute, Blegdamsvej 17, DK-2100 Copenhagen \O{}, Denmark}

\begin{abstract}
We present an asymptotically and unconditionally stable numerical
method to account for the momentum transfer between multiple species.
Momentum is conserved to machine precision. This implies that the
asymptotic equilibrium corresponds to the velocity of the center of
mass.  Aimed at studying dust dynamics, we implement this numerical
method in the publicly available code FARGO3D.  To validate our
implementation, we develop a test suite for an arbitrary number of
species, based on analytical or exact solutions of problems related to
perfect damping, damped sound waves, shocks, local and
global gas-dust radial drift in a disk and, linear streaming instability.  In
particular, we obtain first-order, steady-state solutions for the radial
drift of multiple dust species in protoplanetary disks, in which the
pressure gradient is not necessarily small. We additionally present
non-linear shearing-box simulations of the streaming instability and
compare them with previous results obtained with Lagrangian particles.
We successfully validate our implementation by recovering the
solutions from the test suite to second- and first-order accuracy in
space and time, respectively.  From this, we conclude that our scheme
is suitable, and very robust, to study the self-consistent dynamics of
several fluids.  In particular, it can be used for solving the
collisions between gas and dust in protoplanetary disks, with any
degree of coupling.
\end{abstract}


\section{Introduction}

Protoplanetary disks (PPDs) are composed of a collection of gases and
dust grains orbiting a young star. In general, the dynamics of this
mixture is complex and self-consistent calculations -- including the
coupling between different species -- are required to produce sensible
models. In addition, there is little doubt that self-consistent dust
evolution is necessary to correctly interpret observations of PPDs.

Unperturbed gaseous disks rotate at a sub-Keplerian speed because of
the pressure gradient.  However, dust particles are pressureless, so
they tend to move at the Keplerian speed. This miss-match between the
gas and dust velocities results in a headwind that exchanges momentum
and energy between the species. When considering an unperturbed disk
formed by one gas and one dust species, \cite{Whipple1972} and
\cite{Weidenschilling1977} showed that the dust species drifts inwards
spiraling towards the central star. Momentum conservation makes the
gas rotate slightly faster, but because the dust-to-gas mass ratio is
usually small, gas dynamics is mostly unaffected. However, under some
circumstances, local dust concentrations in the disk can modify the
gas dynamics significantly. For example, dust can accumulate in
vortices in transitional disks \citep[e.g.][]{Barge1995, Lyra2013,
  Zhu2014, Ragusa2017}, in lopsided disks \citep{Baruteau2016} and at
the edges of planet-carved gaps \citep[e.g.][]{Dipierro2017,
  Weber2018}. Dust can also concentrate due to torques exerted by
low-mass planets \citep[e.g.][]{Benitez-Llambay2018, Chen2018}. It has
also been shown that dust can concentrate because of vortices induced
by the self-organization due to the Hall effect in magnetized disks
\citep[e.g.][]{Bethune2016, Krapp2018}, among many others mechanisms.

Because of the importance of dust dynamics in PPDs, several tools have
been developed to study this problem numerically
\citep[e.g.,][]{Johansen2004, Fromang2006b, Paardekooper2006a,
  Youdin2007, Balsara2009, Bai2010, Hanasz2010, Miniati2010,
  Laibe2012a, Laibe2014, Yang2016, Baruteau2016, Price2018, Riols2018,
  Chen2018,Hutchison2018, Stoyanovskaya2018b}.

Two different approaches are usually followed when solving the
dynamics of dust embedded in a gaseous medium. Dust is usually modeled
either as Lagrangian particles or as a pressureless fluid. For
example, Lagrangian particles are used in the public codes PIERNIK
\citep{Hanasz2010, Drcakowska2010}, PENCIL
\cite{Brandenburg2002,Yang2016}, ATHENA \citep[][]{Stone2008,Bai2010},
and PHANTOM \citep{Price2018}. A smaller number of examples can be
found treating dust as a pressureless fluid, for example, PIERNIK
\citep[e.g.][]{Kowalik2013a}, FARGO\_THORIN \citep{Chrenko2017} and
MPI-AMRVAC \citep{MPIAMRVAC}.

To our knowledge, only two implementations are able to solve the
dynamics of multiple fluids: PIERNIK and MPI-AMRVAC. Neither of these
codes are able to exploit the computing power of Graphics Processing
Units (GPUs), which have proven to be an excellent tool for solving
protoplanetary disk-related problems
\citep[e.g.][]{Fung2014,Benitez-Llambay2016}.

In this paper we present a numerical method to solve the momentum
transfer between multiple species in a precise and stable manner. We
show that the implementation of this method in the publicly available
GPU code FARGO3D \citep{Benitez-Llambay2016} correctly describes the
self-consistent dynamics of a mixture of gas and multiple pressureless
dust species.

The goals of this paper are, (i) to comprehensively describe the
numerical method together with its most important properties, (ii) to
develop a test suite for an arbitrary number of fluid species and,
(iii) to validate our implementation in FARGO3D, not only by
recovering the solutions of the test suite but also by studying the
numerical convergence.

This paper is organized as follows: in Section\,\ref{sec:numericalmethod}, we
present and discuss the properties of the numerical method used to
solve the momentum transfer between multiple species and its
implementation in FARGO3D. In Section\,\ref{sec:sec3}, we present a
test suite and compare the obtained analytical or exact solutions with
those resulting from our implementation. In this section, we
additionally show results of the non-linear evolution of the two-fluid
streaming instability. Finally, in Section\,\ref{sec:sec4} we discuss
the main results and perspectives of this work.

\section{Numerical scheme}
\label{sec:numericalmethod}

To present our implementation, we consider a set of $N$ species in
which the temperature depends on the spatial coordinates only. In this
work, we furthermore neglect the possibility of mass transfer between
different species, which can be important in, for instance, dust
coagulation processes or chemical reactions (see
Appendix\,\ref{ap:source_term} for a discussion about this
implementation). We do not consider viscous or external body forces,
whose implementation has been presented in \cite{Benitez-Llambay2016}.
Under these assumptions, the dynamics of the system is completely
described by the continuity and Euler equations for each species,
which contain an additional term accounting for the momentum transfer
between them \citep[e.g.][]{Braginskii1965,Benilov1997}. Labeling the
density and velocity by $\rho$ and ${\bf v}$ respectively, the
equations describing the $N$-species system are
\begin{align}
\frac{D \rho_i}{D t} &= -\rho_i \nabla \cdot {\bf v}_i \,, \label{eq:continuity}\\
\frac{D {\bf v}_i}{D t} &=  - \frac{\nabla P_i}{\rho_i} + \frac{{\bf F}_{i}}{\rho_i}  \label{eq:momenta}\,,
\end{align}
with $i=1,\dotsc,N$ being an index referring to each species, $D/D t =
\partial/\partial_t + {\bf v}\cdot\nabla {\bf v}$ is the material
derivative and $P_i$ is the pressure associated to the $i$-th
species. The drag force per unit volume, ${\bf F}_{i}$, is defined as
\begin{equation}
{\bf F}_{i} = - \rho_i \sum_{j\neq i} \alpha_{ij} \left({\bf v}_i - {\bf v}_j\right)\,,
\end{equation}
with $\alpha_{ij}$ the collision rate between species $i$ and
$j$. This collision rate parameterizes the momentum transfer per unit
time and is, in general, a function of the physical properties of the
species and their mutual relative velocity.
Momentum conservation implies
\begin{equation}
\label{eq:symmetry}
\rho_i \alpha_{ij} = \rho_j \alpha_{ji}\,.
\end{equation}

\subsection{Implicit update}

We solve the $4N$ equations described by \eqref{eq:continuity} and
\eqref{eq:momenta} in the framework of the operator splitting
approximation \citep[e.g.][]{Hawley1984, Stone1992}. In this formalism
each equation is usually split into two, (i) the transport step and
(ii) the source step \citep[see
  e.g.][]{Stone1992,Benitez-Llambay2016}. In our implementation, the
collision terms are solved as a new substep within the source step
(see Section \ref{subsec:implementation} for more details). In this
approximation, the additional equation that needs to be solved
is\footnote{It is worth noticing that the method presented in this
  work holds in the Lagrangian formalism when, in
  Eq.\eqref{eq:new_source}, the partial time derivative is replaced by
  the material one.}
\begin{equation}
\label{eq:new_source}
\frac{\partial {\bf v}_i}{\partial t} =  \frac{{\bf F}_{i}}{\rho_i }\,.
\end{equation}

In FARGO3D, the source step is solved using explicit updates. However,
a very restrictive stability condition appears when solving a mixture
of multiple fluids explicitly \citep[see e.g.][]{Vorobyov2018,
  Stoyanovskaya2018}. In this case, the time step becomes small for
large collision rates. Thus, it is convenient to adopt an implicit
scheme to solve Eq.\,\eqref{eq:new_source}.  The most straightforward
formula is obtained by expressing this equation in finite differences
and evaluating the velocities on the R.H.S. in the advanced time,
i.e.,
\begin{equation}
\label{eq:implicit_form}
\frac{{\bf v}_i^{n+1} - {\bf v}_i^{n}}{\Delta t}  = - \sum_{j\neq i} \alpha_{ij}^n \left({\bf v}_i^{n+1} - {\bf v}_j^{n+1}\right)\,.
\end{equation}
Eq.\,\eqref{eq:implicit_form} corresponds to a set of $3N$ linear
algebraic equations for the unknown ${ \bf v}^{n+1}$ velocities, which
can be written in a more convenient way as
\begin{equation}
\label{eq:linear_system}
{\bf v}_i^{n+1}\left[ 1 + \Delta t \sum_{j\neq i} \alpha_{ij}^n \right] - \Delta t \sum_{j\neq i} \alpha_{ij}^n {\bf v}_j^{n+1}  = {\bf v}_i^n\,.
\end{equation}
A compact form of Eq.\,\eqref{eq:linear_system} is obtained by
defining the column vectors
\begin{equation}
{\bf V}_k = \left[\,{\bf v}_{1}\cdot {{\bf e}_k}\,,\, \dots\,,\, {\bf v}_{N}\cdot {{\bf e}_k}\,\right]^{\mathsf{T}}\,,
\end{equation}
that is, ${\bf V}_k$ is formed by the projection of the velocity of
each species along the direction of the unit vector ${\bf e}_k$, with
$k=1, 2, 3$. The superscript $\mathsf{T}$ stands for transpose. This
makes it possible to write Eq.\,\eqref{eq:linear_system} as the matrix
equation
\begin{equation}
\label{eq:matrix_equation}
{\bf T} {\bf V}^{n+1}_k = {\bf V}^n_k\,,
\end{equation}
where 
\begin{equation}
{\bf T} = {{\bf I} + \Delta t \bf M}\,
\end{equation}
is an $N\times N$ matrix, ${\bf I}$ is the identity matrix, and the
$(i,j)$-element of the matrix ${\bf M}$ is
\begin{equation}
\label{eq:M}
{\bf M}_{ij} \equiv \sum^{N}_{k\neq i}{\alpha^n_{ik}} \delta_{ij} - \alpha^n_{ij} \left(1-\delta_{ij}\right)\,.
\end{equation}
The first and second terms in Eq.\,\eqref{eq:M} set the diagonal and
non-diagonal elements of ${\bf M}$, respectively. Because $\bf T$ is
nonsingular (see Appendix \ref{ap:eigenvalues}), the solution of
Eq.\,\eqref{eq:matrix_equation} exists.

\cite{Stone1997} showed a simple solution of
Eq.\,\eqref{eq:matrix_equation} for two species. However, the
complexity of these solutions rapidly increases with $N$, making them
impractical. In this paper we solve Eq.\,\eqref{eq:matrix_equation}
numerically by means of Gaussian elimination with partial pivoting
\citep[see e.g.][Chapter 2]{Press2007}.

\subsection{Properties of the implicit scheme}

Two important properties of the method arise from
Eq.\,\eqref{eq:matrix_equation}. These are momentum conservation to
machine precision and asymptotic stability for any $\Delta t$.

\subsubsection{Momentum conservation to machine precision} 

The implicit scheme, defined by Eq.\,\eqref{eq:matrix_equation},
conserves total momentum to machine precision. This property can be
demonstrated by comparing the momentum before and after the
application of the operator ${\bf T}$.

We first calculate the momentum of the system at time $t_n$, and
write the old velocities in terms of the new ones via
Eq.\,\eqref{eq:matrix_equation}. Defining $a_{ij} = \Delta t
\alpha^{n}_{ij}$, it follows that
\begin{align}
 &\sum_i \rho_i {\bf v}_i^{n} \nonumber \\
	&= \sum_i \rho_i \sum_j \left[\left( 1 + \sum_{k\neq i} a_{ik} \right)\delta_{ij} - a_{ij} \left(1 - \delta_{ij}\right)\right] {\bf v}_j^{n+1} \nonumber \\
	&= \sum_j  {\bf v}_j^{n+1} \sum_i \rho_i \left[\left( 1 + \sum_{k\neq i} a_{ik} \right)\delta_{ij} - a_{ij} \left(1 - \delta_{ij}\right)\right] \nonumber \\
	&= \sum_j  {\bf v}_j^{n+1} \left[ \rho_j + \rho_j \sum_{k\neq i} a_{jk}  - \sum_{i\neq j} \rho_i  \frac{\rho_j}{\rho_i}a_{ji} \right]  \nonumber\\
	&= \sum_j  {\bf v}_j^{n+1} \rho_j \left[ 1 + \sum_{k\neq i} \left(a_{jk}  - a_{jk}\right) \right]  \nonumber \\
	&= \sum_j  \rho_j {\bf v}_j^{n+1}\,,
\end{align}
where we have used the condition \eqref{eq:symmetry} and replaced $i$
by $k$ in the last step. In this calculation, the densities are
evaluated at time $t_n$ because the collision step does not modify
them. We thus conclude that,
\begin{equation}
\sum_i \rho_i {\bf v}_i^{n+1} = \sum_i \rho_i {\bf v}_i^{n} = \dots = \sum_i \rho_i {\bf v}_i^{0}\,,\\
\end{equation}
implying that the implicit scheme conserves momentum to machine
precision.
\subsubsection{Asymptotic stability}
\label{sec:stability}
The implicit scheme defined by Eq.\,\eqref{eq:matrix_equation} is
asymptotically stable. This is,
\begin{equation}
\lim_{n\rightarrow \infty} {\bf T}^{-n}{\bf V}^{0}_k - {\bf c} = 0\,,
\end{equation}
for some constant vector ${\bf c}$ and any vector ${\bf V}^0_k$. To
prove this property we use that ${{\bf{T}}^{-1}}$ is a right
stochastic and strictly positive matrix (see Appendix
\ref{ap:stochastic} and \ref{ap:positive}). Hence, from the
Perron-Frobenius theorem, ${\bf T}^{-1}$ converges to a matrix with
identical rows, i.e.,
\begin{equation}
\lim_{n\to \infty} \left( {\bf T}^{-n}\right)_{ij} = p_j\,,
\end{equation}
where $p_j$ is the $j^{\rm th}$ element of a vector ${\bf p}$. In
the following, we only use that ${\bf p}$ is constant.

By definition, for any direction ${\bf e}_k$, the implicit scheme
satisfies
\begin{equation}
{\bf V}_k^{n+1} = {\bf T}^{-1} {\bf V}_k^n = \cdots = {\bf T}^{-\left(n+1\right)}{\bf V}_k^0\,.
\end{equation}
It then follows that the asymptotic limit is
\begin{align}
\label{eq:lim1}
\lim_{n\to \infty} {\bf V}_k^{n+1} &= \lim_{n\to \infty} {\bf T}^{-\left(n+1\right)} {\bf V}_k^0 \equiv {V}_{{\rm c},k} {\bf 1}^\mathsf{T}.
\end{align}
where ${\bf 1}^{\mathsf{T}}$ is a vector whose elements are all equal
to one and $V_{{\rm c},k} = {\bf p} \cdot {\bf V}^{0}_{k}$. Since
momentum is conserved, it follows that
\begin{align}
\label{eq:vcm0}
V_{{\rm c}, k}  \sum_{j=1}^{N} \rho_j = \sum_{j=1}^{N}  \rho_j  V_{jk}^0\,,
\end{align}
from which we prove that the asymptotic limit \eqref{eq:lim1}
corresponds to the velocity of the center of mass, ${V}_{{\rm CM},k}$,
defined as
\begin{align}
\label{eq:vcm}
{V}_{{\rm CM}, k}  = \frac{\sum_{j=1}^{N}  \rho_j  V^0_{jk}}{\sum_{j=1}^{N}  \rho_j} \,.
\end{align}
Thus, we conclude that the numerical method converges asymptotically
to the velocity of the center of mass, and this is independent of the
choice of $\Delta t$.

Now, we address the problem of the stability and convergence for any
sufficiently large time step, that is
\begin{equation}
\lim_{\Delta t \to \infty} {\bf T}^{-1} {\bf V}^{n}_{k} - {\bf d} = 0\,,
\end{equation} 
for some constant vector ${\bf d}$ and any vector ${\bf V}^{n}_{k}$.
In Appendix \ref{ap:diagonal}, we show that ${\bf T}^{-1}$ is
diagonalizable, with diagonal form
\begin{equation}
{\bf \Lambda}_{ij} = \frac{1}{1 + \Delta t \lambda_{Mi}} \delta_{ij}\,,
\end{equation}
where $\lambda_{Mi}$ are the eigenvalues of ${\bf M}$, with
$\lambda_{Mj} = 0$ for some $j$. Because ${\bf T}^{-1}$ is right
stochastic, $\lambda_{Mj} = 0$ has algebraic multiplicity equal to
one.
Thus, for any sufficiently large time step, all the entries of ${\bf
  \Lambda}$ approach zero, except $\Lambda_{jj} = 1$.
Then, Eq.\,\eqref{eq:matrix_equation} can be decoupled, and adopts the
form
\begin{equation}
\label{eq:diag}
 \hat{\bf{V}}_{k}^{n+1} = {\bf{\Lambda}} \hat{{\bf{V}}}_{k}^{n}\,,
\end{equation}
where $\hat{\bf{V}}_k = {\bf P}^{-1}{\bf V}_k$, with ${\bf P}$ the
matrix whose columns are the eigenvectors of ${\bf T}^{-1}$.  In the
limit of large $\Delta t$, Eq.\,\eqref{eq:diag} reads
\begin{equation}
\lim_{\Delta t \to \infty} \left(\hat{\bf{V}}_{k}\right)^{n+1}_i = \hat{V}_{j,k}^{n} \delta_{ij}\,.
\end{equation}
Because we set $\Lambda_{jj} = 1$ and ${\bf T}^{-1}$ is right
stochastic, all the entries of the column ${\bf P}_{j}$ are equals to
one, that is, ${\bf P}_j = {\bf 1}^{\mathsf{T}}$.
We thus obtain
\begin{equation}
\label{eq:lim2}
\lim_{\Delta t \to \infty} {\bf V}_{k}^{n+1} =\lim_{\Delta t \to \infty} {\bf P} \hat{\bf V}^{n+1}_{k} = \hat{V}_{j,k}^{n} {\bf P}_{j} = \hat{V}_{j,k}^{n} {\bf 1}^{\mathsf{T}}\,,
\end{equation}
which is equivalent to Eq.\,\eqref{eq:lim1}.

Eqs.\,\eqref{eq:lim1} and \eqref{eq:lim2} allow us to conclude
\begin{equation}
\lim_{n\to \infty} {\bf T}^{-n} {\bf V}^{0}_k = \lim_{\Delta t \to \infty} {\bf T}^{-1} {\bf V}^{n}_k = V_{\rm CM} {\bf 1}^{\mathsf{ T}}\,,
\end{equation}
and the implicit scheme is thus asymptotically and unconditionally stable.

\subsection{Dust as a pressureless fluid}
\label{subsec:dustfluid}

In the case of a system composed of gas and several dust species, dust
can be modeled as a pressureless fluid.  It is clear that this
approximation fails in describing the dynamics of systems dominated by
crossing trajectories, a regime prone to develop when gas and dust
species are coupled very weakly.

In our implementation, we neglect collisions between dust species and
consider only the interaction between the gas and dust fluids. Thus,
dust species interact indirectly between them via their coupling with
the gas.  Referring to the gas species by the index $\rm g$, after
using the condition \eqref{eq:symmetry}, the collision rate can be
written as
\begin{align}
\label{eq:alpha_dust}
\alpha_{ij} \equiv \alpha_i \delta_{j{\rm g}} + \epsilon_j \alpha_j\delta_{{\rm g}i}\,,
\end{align}
with $\epsilon_j = \rho_j/\rho_{\rm g}$ and $\delta_{i{\rm g}}$ the Kronecker delta.  

In the context of PPDs, the collision rate is usually parameterized
via the so-called Stokes number, ${T_{{\rm s}}}$. It is a
dimensionless parameter that characterizes the collision rate in units
of the local Keplerian frequency, $\Omega_{\rm K}$, such that
\begin{equation}
\label{eq:alpha_st}
\alpha_i \equiv \frac{\Omega_{\rm K}}{T_{{\rm s}i}}\,.
\end{equation}
The Stokes number depends on the properties of the gas, the
dust-grains and their relative velocity \citep[see
  e.g.][]{Safronov1972, Whipple1972}.  For simplicity, in this paper,
we assume that the Stokes number does not depend on the velocity of
the fluid, which has been referred in the past as the linear drag
regime \citep[see][]{Laibe2011}.  In what follows, we consider the
Stokes number to be constant. Nevertheless, our implementation remains
valid when the Stokes number is allowed to vary in space. One example
of this is when the dust is characterized by its particle size
\citep[see e.g.][]{Weber2018}. The more general case of a Stokes
number depending on the relative velocity is presented in Appendix
\ref{ap:NL}.

\subsection{Implementation in FARGO3D}
\label{subsec:implementation}

\begin{figure}
  \begin{center}  
    \includegraphics[]{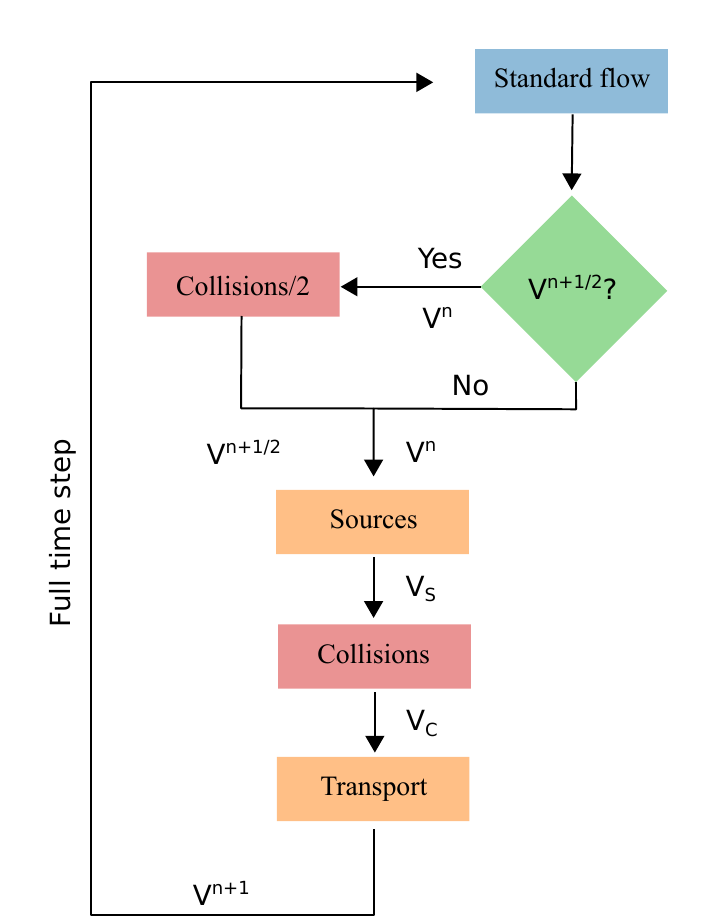}
  \end{center}
  \caption{Flowchart of our implementation. During a generic time step
    $\Delta t$, depending on whether the predictor step is required,
    we call the collision routine using a time step $\Delta t/2$, and
    obtain a partially updated velocity $V^{n+1/2}$. We then update
    the velocities by sources and use the output, $V_{\rm S}$ as input
    for the collision step. After this, we use the updated
    velocities, $V_{\rm C}$, as input for the transport step, from
    which we obtain the updated velocity, $V^{n+1}$. The flow then
    returns to the standard flow, from which a full update has been
    performed.}
  \label{fig:flowchart}
\end{figure}

We now describe the implementation of the implicit scheme in the code
FARGO3D.  We first note that the collision term, described by
Eq.\,\eqref{eq:new_source}, is decoupled from the source and transport
substeps. Thus, we evolve every species according to the same
algorithms described in \cite{Benitez-Llambay2016}.

The implicit scheme for solving the collision term involves an extra
substep, in which the velocity of each species is partially updated
according to the Eq.\,\eqref{eq:matrix_equation}. There are three
different options to place this additional partial update: (i) before
the source step, (ii) after the source step and before the transport
step or, (iii) after the transport step.  Options (i) and (iii) are
equivalent after the first time step, which we discard because the
dust species do not reach the asymptotic limit in the presence of
additional forces. In this case the relative velocity between the dust
and gas asymptotes to its terminal velocity, where the additional
forces are in balance with the drag forces. Since in options (i) and
(iii) the drag forces are computed in the absence of additional
forces, they cannot correctly reproduce this limit
\citep{Booth2015}. Option (ii) reproduces this limit because
evaluating the collision term after the source term is equivalent to a
solution treating both terms together.

While not strictly necessary, the coupling between the source and
collision steps can be improved by adding a predictor step before the
source step. The source step consists of a partial update of the form:
\begin{equation}
\frac{\partial {\bf v}}{\partial t} = \mathcal{S}({\bf v})\,,
\end{equation}
where $\mathcal{S}$ are sources that depend on the velocities. In
finite differences, the previous equation reads:
\begin{equation}
{\bf v}^{n+1} = {\bf v}^{n} + \Delta t \mathcal{S}({\bf v}^{*})\,.
\end{equation}
In the standard implementation, we assume ${\bf v}^* = {\bf
  v}^n$. However, we can improve the coupling between collision and
source steps by setting ${\bf v}^* = {\bf v}^{n+1/2}$, i.e., by
estimating an advanced velocity from the collision step with a time
step $\Delta t/2$. We then compute the source step using a full time
step and finally calculate the collision step with a full time
step. In this paper, we always use the predictor step
when the source terms depend on the velocities. 

For completeness, in Fig.\,\ref{fig:flowchart} we present a flowchart
of our implementation.
During a generic time step, depending on whether the predictor step is
required, we call the collision routine using a time step $\Delta t/2$
and obtain a partially updated velocity.
We then update the velocities of all the species by the standard
source terms and use the updated velocities as input for the
collision step. During this step we solve
Eq.\,\eqref{eq:matrix_equation} and then use the updated velocities as
input for the transport step. After the transport step, a full update
has been performed.

\begin{figure*}[htb!]
  \centering
  \includegraphics[]{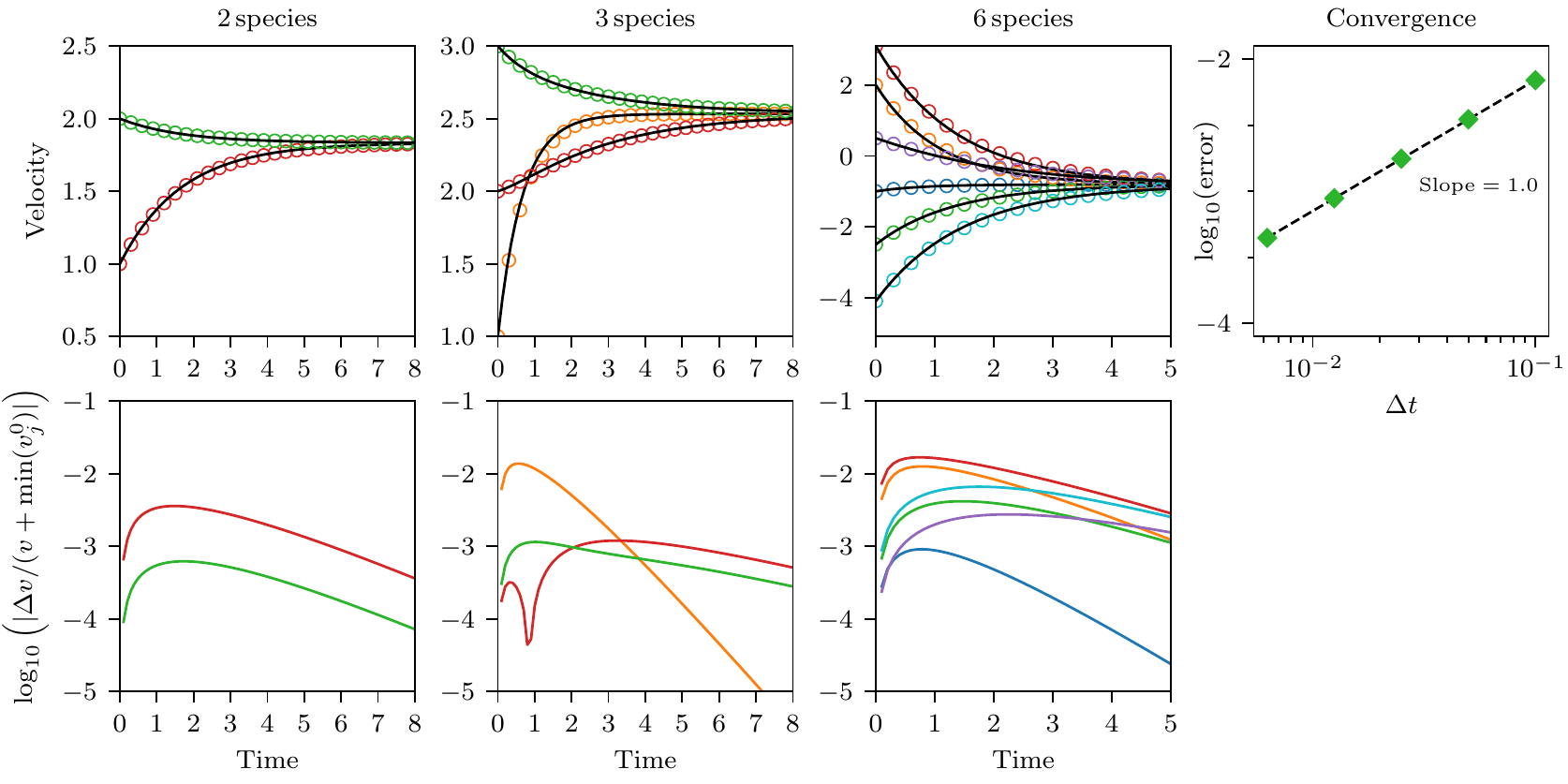}
  \caption{Upper panels: Time evolution of the velocity for the
    various configurations described in Table \ref{tab:damping}. From
    left to right, we plot the evolution of two, three, and six
    species. The solid lines correspond to the analytical solution,
    given by Eq.\,\eqref{eq:solution_damping}. The open circles were
    obtained with our implementation, for which each color represents
    a different species. In all the panels, the velocities converge to
    the velocity of the center of mass of the system. Lower panels:
    Time evolution of the relative error between the numerical and the
    analytical solutions. The color code is the same one for the upper
    and lower panels. The time evolution of the velocities and errors
    are shown for the run with $\Delta t = 0.1$. In the rightmost
    panel, we plot, for the case with six species, the error (see
    Eq.\,\eqref{eq:error_1}) as a function of the time step $\Delta
    t$, for five different time steps.}
  \label{fig:damping}
\end{figure*}

\section{Numerical tests}
\label{sec:sec3}

In this section, we present a test suite considering multiple dust
species.
We use these tests to validate the accuracy, convergence properties
and robustness of the method and implementation described in
Section\,\ref{sec:numericalmethod}. In all the following tests, the
numerical solutions were obtained using a Courant-Friedrichs-Lewy
(CFL) factor of 0.44 \citep{Benitez-Llambay2016}, unless a different
value is specified.

\subsection{Time evolution of a set of colliding species}
\label{sec:damping}

When a set of $N$ species evolves under the sole effect of the
collision term, simple asymptotically convergent analytical solutions
can be found. This simple test problem validates the correct
implementation of the matrix solver and, at the same time, illustrates
the two properties described in the previous section.

As a first step, we show the steady-state solution of the problem,
which gives insight into the fundamental property of the physical
system, that is, the convergence of all the velocities towards the
velocity of the center of mass. In Section\,\ref{sec:stability}, we
have already shown that the implicit scheme satisfies this condition.

The steady-state momentum equation for a set of $N$ species when
considering only the drag force, reduces to the matrix equation
\begin{equation}
  \label{eq:homo}
        {\bf M} {\bf V}_k = 0\,.
\end{equation}
Because the matrix ${\bf M}$, defined by Eq.\,\eqref{eq:M}, is
singular (see Appendix \ref{ap:eigenvalues}), the system
\eqref{eq:homo} admits a non-trivial solution.
By direct calculation it can be shown that the $(i,j)$-element of the
echelon form of ${\bf M}$ is
\begin{equation}
  \label{eq:echelon}
  E_{M,ij} = \delta_{ij} - \delta_{jN}\,,
\end{equation}
This observation, combined with momentum conservation, allows us to
conclude that
\begin{equation}
  {V}_{1k} = \dots = {V}_{Nk} = {{V}}_{{\rm CM},k}\,,
\end{equation}
where ${{V}}_{{\rm CM},k}$ is the velocity of the center of mass,
given by Eq.\,\eqref{eq:vcm}.

\subsubsection{Evolution towards steady-state}

\label{sec:evol_damp}

We are not only interested in the steady-state solution of the system
but also in the time evolution towards this asymptotic
steady-state. The problem is equally described for any component of
the velocity, so it is effectively a collection of 1D problems. Hence
we omit the subscript $k$.

The temporal evolution of the system is described by the solution of
\begin{equation}
  \label{eq:time_evolution}
  \frac{\partial {\bf V}}{\partial t} + {\bf M} {\bf V} = 0\,.
\end{equation}
Without loss of generality, by expressing the solution of
\eqref{eq:time_evolution} as ${\bf V}(t) = \sum_j \tilde{\bf V}_{j}
e^{-\lambda_j t}$, we reduce Eq.\,\eqref{eq:time_evolution} to the
eigenvalue problem
\begin{equation}
  {\bf M} \tilde{\bf V}_j = \lambda_j \tilde{\bf V}_j\,.
\end{equation}
For simplicity, we define the collision rate $\alpha_{ij} \equiv
\alpha_0$ for $i>j$ and $\alpha_{ij} = \rho_j/\rho_i \alpha_0$ for
$i<j$, such that condition \eqref{eq:symmetry} is satisfied.

Defining the function
\begin{equation}
  \zeta_j = \sum_{m=j+1}^N \rho_m\,,
\end{equation}
the eigenvalues of ${\bf M}$ adopt the expression
\begin{equation}
\begin{aligned}
  \lambda_{j<N} &= \alpha_0\left( j + \frac{\zeta_j}{\rho_j}\right)\,, \\
  \lambda_N &= 0\,,
\end{aligned}
\end{equation}
with associated eigenvectors
\begin{equation}
  \begin{aligned}
  \tilde{{\bf V}}_{j<N} &= {\bf e}_j - \frac{1}{\zeta_j}\sum_{m={j+1}}^{N-1} \rho_{m-1} {\bf e}_{m} \,,\\
  \tilde{{\bf V}}_N &= \sum_{k=1}^N {\bf e}_k\,.
\end{aligned}
\end{equation}

Thus, the solution reads
\begin{equation}
  \label{eq:solution_damping}
  \begin{aligned}
    v_{j<N}(t) &= -\sum_{k=1}^{j-1} \frac{\rho_k c_k }{\zeta_k} e^{-\lambda_k t} + c_j e^{-\lambda_j t} + c_N\,, \\ 
    v_N(t) &= -\sum_{k=1}^{N-1} \frac{\rho_k c_k }{\zeta_k} e^{-\lambda_k t} + c_N\,,
  \end{aligned}
\end{equation}
where the coefficients $c_j$ are
\begin{equation}
  \begin{aligned}
  c_{j<N} &= {v}_j^0 - V_{\rm CM} + \sum_{k=1}^{j-1}\frac{\rho_k c_k}{\zeta_k}\,,\\
  c_{N} &= V_{\rm CM}\,. 
\end{aligned}
\end{equation}

\subsubsection{Numerical solution}

\begin{table}[b!]
  \begin{center}
    \caption{Initial density, velocity, eigenvalues and coefficients needed to compute the solution \eqref{eq:solution_damping}. In all the cases we set $\alpha_0 = 10^{-1}$.}
    \label{tab:damping}
    \begin{tabular}{ccccc}
      \decimals
      \hline\hline
      $j$ & $\rho_j$ & $v_j^0$ & $\lambda_j$ & $c_j$ \\
      \hline
      Two fluids \\
      1 & 0.2 & 1.0  & 0.6000000 & -0.8300000  \\
      2 & 1.0 & 2.0  & 0.0000000 &  1.8333333  \\
      \hline
      Three fluids \\
      1  & 0.2 & 1.0  & 1.5000000 & -1.5333333  \\
      2  & 1.0 & 2.0  & 0.3800000 & -0.6428571  \\
      3  & 1.8 & 3.0  & 0.0000000 &  2.5333333  \\
      \hline
      Six fluids \\
      1 & 1.0 & -1.0 & 1.3500000 & -0.1925926  \\
      2 & 1.5 &  2.0 & 0.9333333 &  2.7920000  \\
      3 & 2.0 &  3.1 & 0.7500000 &  4.2727273  \\
      4 & 2.5 & -2.5 & 0.6600000 & -0.3777778  \\
      5 & 3.0 &  0.5 & 0.6166667 &  2.4769231  \\
      6 & 3.5 & -4.1 & 0.0000000 & -0.8074074  \\
      \hline
    \end{tabular}
  \end{center}
\end{table}

\begin{figure*}[htb!]
  \centering
  \includegraphics[]{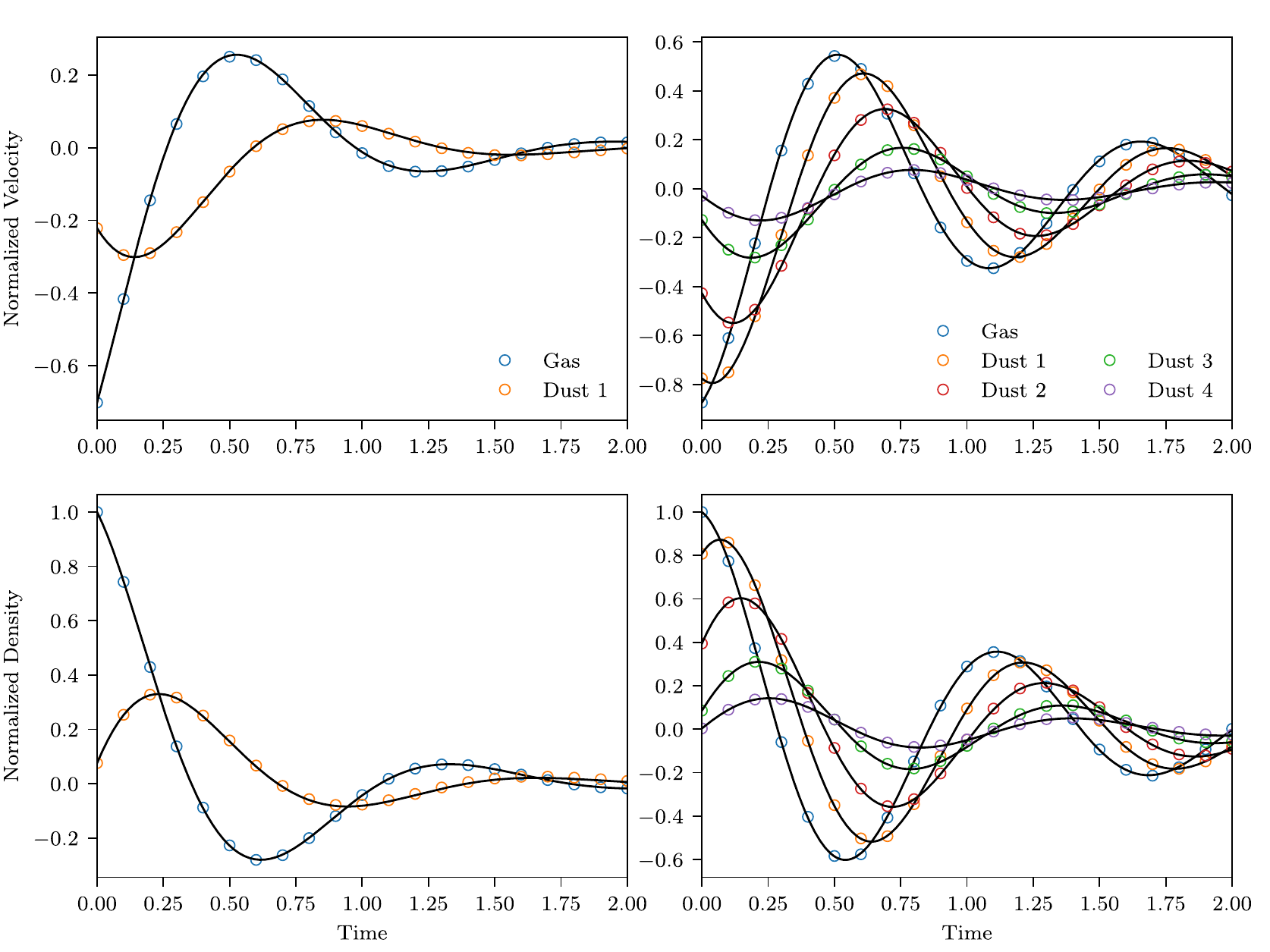}
  \caption{Numerical (open circles) and analytical (solid lines)
    solutions of the test described in Section \ref{sec:sound-wave}
    for the configurations listed in Table \ref{tab:dampwave}. We plot
    the time evolution of the normalized velocity (upper panels) and
    density (lower panels). The results for one and four dust species
    are shown in the left and right panels, respectively. The blue
    circles correspond to the gas, while the other colors correspond
    to the dust species. All the solutions were obtained at $x=0$. The
    normalized density and velocity are defined as $\delta
    \hat{\rho}/(A\rho^0)$ and $\delta\hat{v}/(Ac_{\rm s})$, respectively, with $A =
    10^{-4}$. }
  \label{fig:dampwave}
\end{figure*}

We compare the analytical solution found in the
Sec\,\ref{sec:evol_damp} with that obtained by solving the problem
using the implicit scheme. We study the problem with two, three, and
six different species. In order to do this, we set the initial
condition on a 1D grid with 16 evenly distributed cells, over a
periodic domain. We note, however, that this choice is arbitrary and,
in practice, irrelevant. This is because the solution does not depend
on spatial coordinates. The initial density, velocity, eigenvalues,
and coefficients needed to compute both the numerical and analytical
solutions for each run are summarized in Table \ref{tab:damping}. In
all the cases, we set the collision rate $\alpha_0 = 10^{-1}$.

In the first three panels of Fig.\,\ref{fig:damping} (from left to
right) we plot, for all the species, the time evolution of the
velocities. Each panel corresponds to the different configurations
listed in Table \ref{tab:damping}. In the lower panels of
Fig.\,\ref{fig:damping}, we present the relative error of the velocity
for all the species, for each of the three cases. The time evolution
of the error shows the asymptotic convergence demonstrated in Section
\ref{sec:numericalmethod}.  Independently of the number of species,
there is an excellent agreement between the analytical and numerical
solutions.

\subsubsection{Convergence with time step}

We additionally check, for the case of six fluids, the expected
first-order convergence rate in time of the implicit scheme.
For this test, we performed five identical runs, in which we
progressively decreased the time step by factors of 2, starting with
$\Delta t = 0.1$. In the rightmost panel of Fig.\,\ref{fig:damping},
we plot the error as a function of the time step, defined as
\begin{equation}
\label{eq:error_1}
\textrm{error}(\Delta t) = \left(\sum_{j=1}^N \langle v_j^{\Delta t}(t) - v_j(t)\rangle^2\right)^{1/2}\,,
\end{equation}
where $\langle \rangle$ denotes the time-average.
As expected, the convergence is consistent with a first-order method,
i.e., linear convergence with a slope equal to one.

\subsection{Damping of a sound wave}
\label{sec:sound-wave}

Sound waves are a natural outcome of the fluid equations when pressure
perturbations are considered. Dust fluids, however, cannot support
sound waves. In systems composed of gas and dust species, sound waves
can propagate -- supported by the gas component -- but their
properties are modified due to the coupling to the gas.

Solutions for the case of one gas and one dust species were found by
\cite{Laibe2012a}, who show that sound waves are damped by the effect
of the mutual collision. Solving this problem is relevant since it
provides a direct -- and perhaps the simplest -- way to test the
coupling between the implicit solver and the transport and source
steps.

In this paper, we derive the dispersion relation for the general
problem of one gas and $N-1$ dust species which, together with the
general expression for the eigenvectors, allow us to find the full
solution to the problem.

\bigskip

\bigskip

\subsubsection{Dispersion relation and eigenvectors}

\begin{table*}
  \begin{center}
    \caption{Initial conditions for the damping of the sound wave test.}
    \label{tab:dampwave}
    \begin{tabular}{cccccc}
      \decimals
      \hline\hline
      j & $\rho_j$ & $\delta \hat{\rho}_{j}$ & $\delta \hat{v}_j$ & $t_{{\rm s}j}$ & $\omega$ \\
      \hline
      Two species \\
      g & 1.000000  & 1.000000  & $-0.701960-0.304924i$ & -- & $1.915896 -4.410541i$\\
      1   & 2.240000 & $0.165251-1.247801i$  & $-0.221645+0.368534i$	& 0.4 & \\
      \hline
      Five species &  \\
      g & 1.000000 & 1.000000 & $-0.874365-0.145215i$ & -- & $0.912414 -5.493800i$ \\
      1 & 0.100000 &  $0.080588-0.048719i$ & $-0.775380+0.308952i$ &  0.100000  & \\
      2 & 0.233333 &  $0.091607-0.134955i$ & $-0.427268+0.448704i$ &  0.215443 &  \\
      3 & 0.366667 & $0.030927-0.136799i$ & $-0.127928+0.313967i$ &   0.464159 & \\
      4 & 0.500000 &  $0.001451-0.090989i$ & $-0.028963+0.158693i$ &  1.000000 & \\
      \hline
    \end{tabular}
  \end{center}
\end{table*}

We now derive the dispersion relation for the case of one gas and
$N-1$ dust species and find the general eigenvectors of the problem.
For that, we assume that the gas pressure is given by $P = c_{\rm
  s}^2\rho_{\rm g}$, with a constant sound speed, $c_{\rm s}$, and
define the collision rate between the gas and dust species following
Eq.\,\eqref{eq:alpha_dust}, with $\alpha_j = t_{{\rm s}j}^{-1}$, where $t_{{\rm s}j}$ is
the stopping time.

Assuming solutions of the form $\rho_j = \rho_{j}^0 + \delta \rho_j$
and $v_j = \delta v_j$, with $\rho_j^0$ constant, and neglecting
quadratic terms in the perturbations, the continuity and momentum
equations for the gas and dust species become
\begin{align}
\label{eq:linear}
\frac{\partial \delta \rho_{\rm g} }{\partial t}+ \rho_{\rm g}^0\frac{\partial \delta v_{ {\rm g} }}{\partial x}  &= 0\,, \\
\frac{\partial \delta \rho_{j}}{\partial t}  +  \rho_{j}^0 \frac{\partial \delta v_{j}}{\partial x} &= 0\,, \\
\frac{\partial \delta v_{{\rm g}}}{\partial t} + \sum_{m=1}^{N-1} \frac{\epsilon_m^0}{t_{{\rm s}m}} (\delta v_{\rm g}-\delta v_{m}) + \frac{c_{\rm s}^2}{\rho_{\rm g}^0} \frac{\partial \delta \rho_{\rm g} }{\partial x}  &= 0\,, \\
\frac{\partial  \delta v_{j} }{\partial t} + \frac{1}{t_{{\rm s}j}} ( \delta v_{j} - \delta v_{\rm g})  &= 0, \label{eq:linear_end}
\end{align}
where $j= 1, ..., N-1$ is the index of the dust species.

We first note that the momentum equation is decoupled from the
continuity equation for dust species, so the order of the problem is
effectively reduced from $2N$ to $N+1$. Without loss of generality, we
write any perturbation $\delta f$ as $\delta f = \delta
\hat{f} e^{ikx - \omega t}$, with $k$ a real wavenumber. Thus,
from Eqs.\,\eqref{eq:linear}-\eqref{eq:linear_end}, we obtain the
dispersion relation
\begin{equation}
  \label{eq:dispersion}
  F(\omega,\omega_{\rm s}) \equiv \omega^2\left( 1 + \sum_{m=1}^{N-1} \frac{\epsilon_m}{1-\omega t_{{\rm s}m} }\right) + \omega_{\rm s}^2 = 0\,,
\end{equation}
with $\omega_{\rm s} = k c_{\rm s}$. The singular values $\omega_m = t_{{\rm s}m}^{-1}$ correspond to $\delta v_{\rm g} = 0$, $\delta \rho_{\rm g} = 0$, so are not considered.

Finally, the components of the associated eigenvectors are
\begin{align}
\label{eq:eigen_damp}
\frac{\delta \hat{v}_{{\rm g}}}{c_{\rm s}}  &= -i\frac{\omega}{\omega_{\rm s}} \frac{\delta \hat{\rho}_{{\rm g}}}{\rho^0_{\rm g}} \,,\\
\frac{\delta \hat{v}_{j}}{{c_{\rm s}}} &=  -i\frac{\omega }{\omega_{\rm s}}\frac{1}{(1-\omega t_{{\rm s}j})} \frac{\delta \hat{\rho}_{{\rm g}}}{\rho^0_{\rm g}}\,, \\
\frac{\delta \hat{\rho}_{j}}{\rho^0_{j}}  &=  \frac{1}{1-\omega t_{{\rm s}j}} \frac{\delta \hat{\rho}_{{\rm g}}}{\rho^0_{\rm g}}\,,
\label{eq:eigen_damp_last}
\end{align}
for any $\delta \hat{\rho}_{{\rm g}}$, which completes the solution
of the problem.

Eq.\,\eqref{eq:dispersion} can be written as a polynomial equation of
degree $N+1$. In Appendix \ref{ap:dispersion-soundwave} we show that
at least $N-1$ roots of \eqref{eq:dispersion} are real and positive
and are thus associated with pure damping.  We furthermore identify
the intervals in which they can be found. This allows a simple
bisection algorithm to be used to find them. We additionally explain
how to use Vieta's formulae to find the final two roots which are, in
general, complex. These two complex roots are the most interesting
ones since they describe the propagation of damped sound waves.

\subsubsection{Numerical solution}

We obtain numerical solutions for the oscillatory damped modes. From
the two possible oscillatory modes, we choose only one because the
other is the complex conjugate, producing the same solution but
propagating in the opposite direction. We do not consider the
solutions that correspond to perfect damping because they behave as
those studied in Section \ref{sec:damping}.

We study the cases of one gas fluid combined with one and four dust
species, respectively. As initial condition we set a zero background
velocity, constant background density, $\rho^0_j$, and perturbations,
$\delta f$, of the form
\begin{equation}
  \delta f = A \left[\operatorname{Re}\left(\delta \hat{f}\right) \cos(kx) - \operatorname{Im}\left(\delta \hat{f}\right) \sin(kx)\right]\,,
\end{equation}
where $A$ is a small amplitude needed to ensure linearity. We set its
value to $10^{-4} c_{\rm s}$ and $10^{-4}\rho_{\rm g}^0$ for the
velocity and density perturbations, respectively. We adopt $c_{\rm s}
= 1$. The background densities, perturbation amplitudes, stopping
times, and complex eigenvalue for each case are listed in Table
\ref{tab:dampwave}. We consider a domain of size $L=1$, with spatial
coordinate $x\in\left[0,L\right]$, split into $10^3$ evenly spaced
grid cells. We consider the wavenumber $k=2\pi/L$ and set periodic
boundary conditions.

In Fig.\,\ref{fig:dampwave}, we plot, for the two configurations
considered, the analytical (solid lines) and numerical (open circles)
solutions, measured at $x=0$. The solution corresponding to each
species is plotted with a different color. The first column shows the
solution obtained for one gas and one dust species, while the second
one shows the same for the case of five, one gas and four dust,
species. In the upper and lower panels we plot the normalized
velocity, defined as $\delta \hat{v}/(c_{\rm s}A)$ and the normalized
density, defined as $\delta \hat{\rho}/(\rho A)$, respectively. From
Fig.\,\ref{fig:dampwave}, it is clear that the analytical solution is
successfully recovered by our implementation. This test validates the
coupling of the drag force in combination with the source and
transport steps for a wide range fo stopping times.

\begin{figure}[]
	\centering
	\includegraphics[]{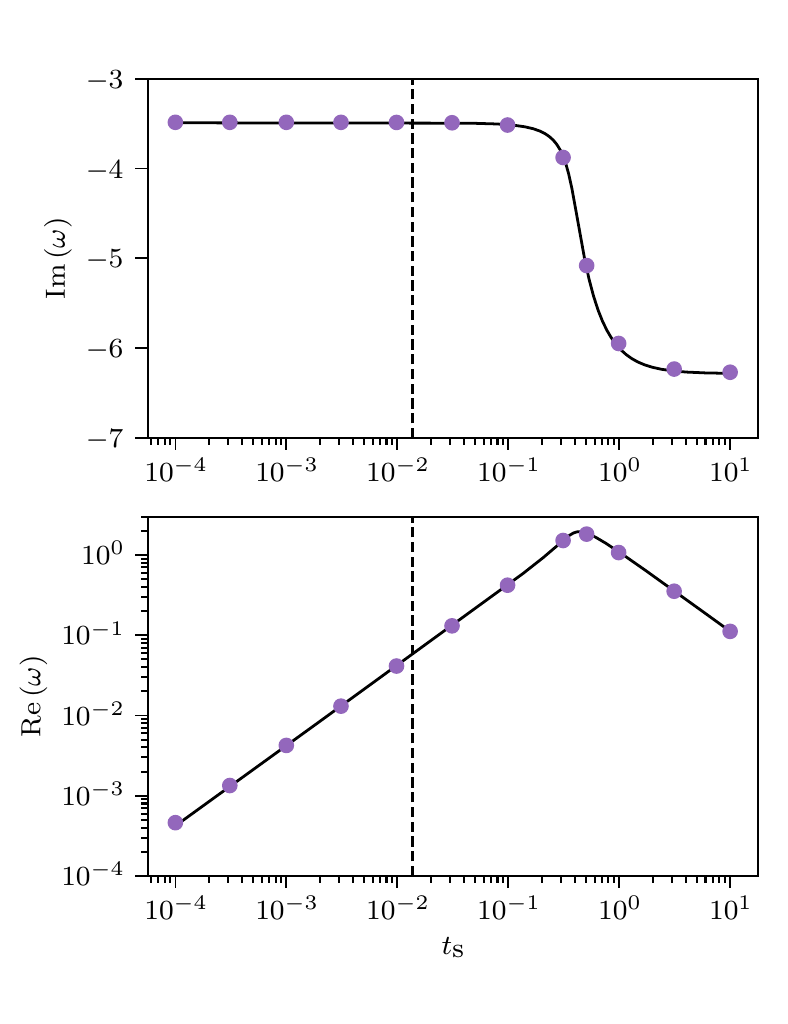}
	\caption{Numerical (circles) and analytical (solid lines)
          imaginary and real part of the eigenvalue $\omega$, as a
          function of the stopping time $t_{\rm s}$, obtained for the
          two-fluid case described in Table \ref{tab:dampwave}. The
          dashed line corresponds to the time step $\Delta t = 1.375
          \times 10^{-2}$, which is fixed for all the runs. }
	\label{fig:dampwave_fit}
\end{figure}

To study the coupling of the implicit scheme with the transport and
source steps in a more challenging situation, we study the damping of
a sound wave for a range of stopping times $10^{-4} \leq t_{\rm s}
\leq 10$ and a fixed time step, such that we test both stiff and
non-stiff regimes for the collisions. We consider the two-fluid
problem described in Table \ref{tab:dampwave}, for different stopping
times. We use a domain of size $L=1$ and 32 cells, which sets a time
step $\Delta t = 1.375 \times 10^{-2}$, given by the standard CFL
condition of FARGO3D. For stopping times smaller than the time step
the regime becomes more and more stiff. Note that, because of the CFL
condition, the degree of stiffness depends on the resolution.  We
integrate the system until it reaches a final time $t=10$. We measure
the damping rate, $\textrm{Re}(\omega)$, and the oscillatory
frequency, $\textrm{Im}(\omega)$ by fitting the numerical
solutions. In the upper and lower panels of
Fig.\,\ref{fig:dampwave_fit}, we show the analytical frequency and
damping rate, respectively, together with the measurements from our
simulations.

Since the error of the implicit scheme converges to zero
asymptotically with $\Delta t$ (see Section
\ref{sec:numericalmethod}), for a fixed time step, the smaller the
stopping time is, the stiffer the regime is and the faster the errors
are damped. Furthermore, the excellent agreement of the oscillatory
frequency allows us to conclude that no phase-error is introduced by
the implicit scheme, in the operator splitting approximation.

\subsection{Shock solution under the presence of dust}
\label{sec:shock}

\begin{figure*}[htb!]
  \centering
  \includegraphics[scale=0.94]{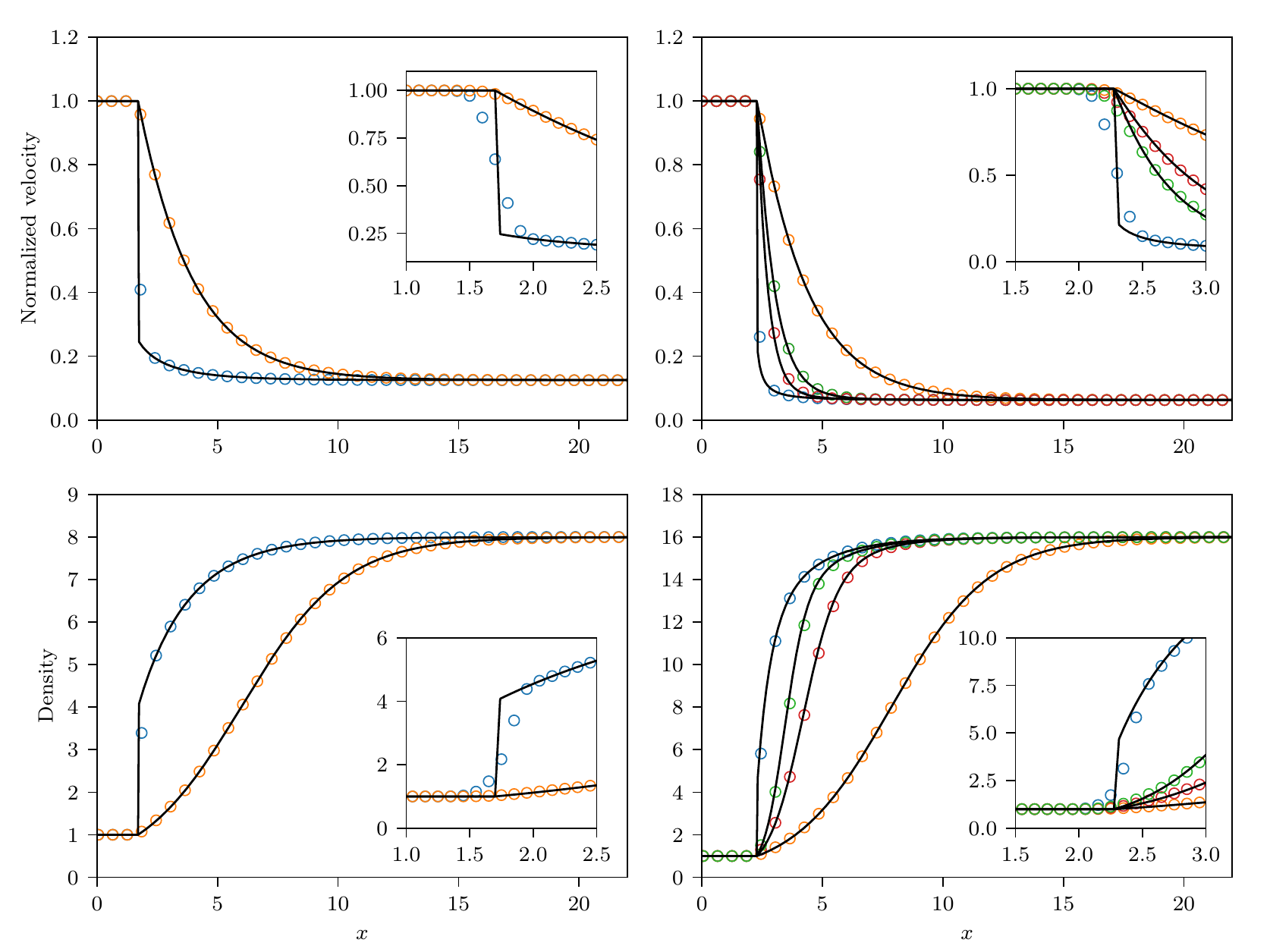}
  \caption{\label{fig:dustyshock} Numerical (open circles) and
    analytical (solid lines) solutions for the shock test problem,
    described in Section \ref{sec:shock}, when considering two (left
    panels) and four (right panels) species. The numerical solution
    was obtained at time $t=500$, starting from an initial jump
    condition. The upper panels show the normalized velocities
    $\omega_{\rm g},{\omega}_i$ of the gas (blue) and dust (orange,
    red and green) species, respectively. The lower panels show the
    density of the gas and dust species, sharing the same color
    code. In the large panels, to allow assessing the quality of the
    asymptotic behavior far away from the shock, the sampling rate for
    the open circles was reduced to 1:6 of the original data. We
    additionally plot, inside each panel, a zoomed region within the
    shocks showing the full sampling, i.e., the open circles
    correspond to the actual grid points. The code resolves the shock
    with 3-4 cells, even when an increasing number of fluids is
    considered. The overall agreement between the numerical and
    analytical solutions is excellent.}
\end{figure*}

\cite{Lehmann2016} found steady-state shock solutions for a mixture of
gas and one dust species. In this paper, we extend one of those
solutions to consider an arbitrary number of dust species, which
provides a simple and effective way to test the response of the dust
species under the presence of a shock in the gas component. This
generalization allows us to test the collision module in combination
with the hydro solver in a challenging regime, in which a steady-state
solution must be achieved but is not, in principle, numerically
guaranteed. Additionally, this test problem allows us to measure how
many cells the shock spreads over in the multiple species
configuration.

\subsubsection{Generalized shock solution for gas and $N$-dust species}

The shock solution is obtained after solving the steady-state
continuity and momentum equations for the gas and $N$-dust species:
\begin{align}
  \frac{\partial}{\partial x}\left(\rho_{\rm g} v_{\rm g}\right) &= 0 \label{eq:dustyshock-contgas}\,,\\
  \frac{\partial}{\partial x}\left({\rho}_i {v}_i\right) &= 0\,, \label{eq:dustyshock-contdust} \\
  \frac{\partial}{\partial x}\left[\rho_{\rm g} \left(v_{\rm g}^2 + c_{\rm s}^2\right)\right] &= -\sum_{i=1}^N K_i \left(v_{\rm g}-{v}_i\right)\,, \label{eq:dustyshock-momgas}\\
  \frac{\partial}{\partial x}\left({\rho}_i {v}_i^2 \right) &= -K_i \left({v}_i-v_{\rm g}\right) \label{eq:dustyshock-momdust}\,,
\end{align}
for $i=1,\dots,N$. $c_{\rm s}$ is the sound speed and, for
simplicity, the collision coefficients $K_i \equiv \rho_g \alpha_{{\rm
    g}i} = {\rho}_i \alpha_{i{\rm g}}$ are assumed to be constant.
After integrating Eqs.\,\eqref{eq:dustyshock-contdust}, we obtain
\begin{equation}
  {\rho}_i {v}_i = {\rho}_{i0} {v}_{i0}\,,
  \label{eq:dustyshock_fistcondition}
\end{equation}
where ${\rho}_{i0} \equiv {\rho}_{i}(x_0)$ and ${v}_{i0} \equiv v_{i}(x_0)$, with $x_0$ an arbitrary
coordinate. Defining the velocity $v_{\rm s} \equiv v_{\rm g0} = {{v}_1}_0 = \dots
= {v_{N}}_0 $, and ${\omega}_i = {v}_i/v_{\rm s}$,
${\omega_{\rm g}} = {v_{\rm g}}/v_{\rm s}$,
Eqs.\,\eqref{eq:dustyshock-momdust} and
\eqref{eq:dustyshock_fistcondition} lead to the following set of
differential equations for the dust velocities
\begin{equation}
  \frac{d{\omega}_i}{d x} = \frac{K_i}{ {\rho}_{i0} {v}_{i0}} \left(\omega_g - {\omega}_{i}\right)\,.
  \label{eq:dustyshock_wd}
\end{equation}
The normalized gas velocity is obtained after integrating the
sum of Eqs.\,\eqref{eq:dustyshock-momgas} and
\eqref{eq:dustyshock-momdust}. This allows $\omega_g$ to be obtained
as the root of the quadratic equation
\begin{equation}
  \omega_\text{g}^2 + \omega_\text{g}\left[ \sum_{i=1}^N \epsilon_i \left({\omega}_i-1\right) - \mathcal{M}^{-2} - 1\right] + \mathcal{M}^{-2} = 0,
  \label{eq:dustyshock_wg}
\end{equation}
where $\epsilon_i = {\rho}_{i0}/\rho_{\rm g0}$ is the dust-to-gas mass
ratio of each species and $\mathcal{M} = v_{\rm s}/c_{\rm s}$ is the Mach
number. Eqs.\,\eqref{eq:dustyshock_wd}, together with the closed
expression for $\omega_g$ given by the solution of
\eqref{eq:dustyshock_wg}, allow us to find the steady-state normalized
velocities as the solution of an initial value problem, described by a
set of $N$ coupled first-order differential equations, with the
initial condition at $x=x_0$.

We use Eq.\,\eqref{eq:dustyshock_fistcondition} and its equivalent for
the gas component to obtain the steady-state density of every
species. Because of the drag force, the velocities are asymptotically
equal far away from the shock, allowing the asymptotic right state
($+$) to be found in terms of the left state ($-$). Defining the left
state as
\begin{equation}
\begin{aligned}
{\rho^{-}_{\textrm{g}}} &= \rho_{\rm g0}\,, \\
{\rho^{-}_{i}} &= \epsilon_i \rho_{\rm g0}\,,\\
\omega_{\rm g}^{-} &= {\omega}_i^{-} = 1.0\,,
\end{aligned}
\end{equation}
the asymptotic right states are
\begin{equation}
\begin{aligned}
\rho_{\textrm{g}/i}^+ &= \frac{\rho_{\textrm{g}/i}^{-}\omega_{\textrm{g}/i}^{-}}{\omega_{\textrm{g}/i}^{+}}\,. \\
\omega_{\textrm{g}}^{+} &= \omega_{i}^{+} = \left(1+\sum_{i=1}^N \epsilon_i\right)^{-1}\mathcal{M}^{-2}\,, 
\end{aligned}
\label{eq:dustyshock_jump}
\end{equation}

\subsubsection{Numerical solution}

\begin{table}[b!]
  \caption{Parameters for the dusty shock test.}
  \begin{center}
    \begin{tabular}{cccccccc}
      \tablewidth{0pt}
      \hline\hline
      Fluids  & $K_1$ & $K_2$ & $K_3$ & $\rho^{-}$ & $\rho^{+}$ & $\omega^{-}$ & $\omega^{+}$ \\
      \hline
      2 & 1.0 & -   & -   & 1.0 &  8.0  & 1.0 & 0.125  \\
      4 & 1.0 & 3.0 & 5.0 & 1.0 &  16.0 & 1.0 & 0.0625\\
      \hline
    \end{tabular}
    \label{tab:dustyshock}
  \end{center}
  \tablecomments{For the two cases we define $\mathcal{M} = 2$, $c_s = 1$ and $\epsilon=1$ for every dust species.}
\end{table}

For this test, we solve the cases of one gas combined with one and
three dust species, respectively. In Table \ref{tab:dustyshock} we
summarize the parameters used for each configuration.

We first obtain what we call the exact solution, given by the solution
of the initial value problem (integrated numerically) described by
Eqs.\,\eqref{eq:dustyshock_wd}-\eqref{eq:dustyshock_wg}, using the
left state of the shock as initial condition. We then obtain the
numerical solution that results from our implementation.

For the numerical solution, we initialize a discontinuity for both the
density and velocity of each fluid. The left and right states are set
equal to the asymptotic right steady-state as obtained from
\eqref{eq:dustyshock_jump}, and given in Table
\ref{tab:dustyshock}. We set the sound speed $c_s = 1$ and the Mach
number $\mathcal{M} = 2$, implying $v_{\rm s} = 2$. The dust-to-gas
mass ratio for all the dust species is set to $\epsilon=1$, so the
left states are all equal. The numerical domain spans from $x=0$ to
$x=40$, sampled over 400 evenly spaced grid points and the initial
jump condition occurs at $x=4$. We use zero gradient boundary
conditions.

In Fig.\,\ref{fig:dustyshock}, we plot the exact (solid lines) and
numerical (open circles) solution for both the two (left panels) and
four (right panels) fluids shock tests. We plot the normalized
velocity and density for both the gas and dust fluids in the upper and
lower panels respectively. Different colors correspond to different
species. To compare the numerical solution with the exact one, we
shifted the exact solution to the shock position. The numerical
solution corresponds to a snapshot taken at $t=500$, a time that is
long enough that no significant variation of the numerical solution is
observed.  The open circles in the main panels are a sub-sampling
(1:6) of the grid points. Inside of each panel, we plot a zoomed-in
region containing the discontinuity to show the quality of the
numerical solution across the shock in the actual grid.

This test indicates that, as expected, the code resolves a shock
within four cells \citep[c.f.][]{Benitez-Llambay2016}. It furthermore
shows that our implementation is able to recover the correct solution
across the shock for all the gas and dust species. The agreement
between the exact and numerical solutions is excellent, successfully
demonstrating the ability of the code to correctly resolve the shock
dynamics.

\subsection{Steady-state, first-order disk-drift solutions}

In this section, we test the coupling between the collision step and
the source and transport steps in our numerical scheme. To accomplish
this, we first find the steady-state radial drift solution for an
arbitrary number of species, to first-order in the velocities with
respect to an exact background. We then compare this analytical
solution with the numerical one.  The background is obtained by
considering pressure gradients (which are not necessarily small) and
neglecting drag forces between species.  This solution generalizes
that obtained by \cite{Nakagawa1986}, who presented a self-consistent
first-order solution with respect to a Keplerian background for a
disk, composed of gas- and one dust-species. In their approach, the
background flow is obtained as the solution of the vertically
integrated disk-equations when neglecting pressure and drag
forces. This assumption implies that both the radial pressure gradient
and the drag force are small perturbations that can be added linearly
to the Keplerian velocity. However, the assumption of a small pressure
gradient is not strictly necessary to find a background solution.

This generalization provides us with improved steady-state disk models
which allow us to thoroughly test our numerical method.

\subsubsection{Generalized steady-state drift solutions}

\label{sec:global_disk_solutions}

In order to find the background solution for the radial drift problem
we work with the vertically integrated disk equations with an
isothermal equation of state, in which the pressure $P = c_{\rm s}^2
\Sigma$, with $c_{\rm s}$ the sound speed and $\Sigma$ the surface
density.
Defining the aspect-ratio $h = c_s(r)/v_{\rm K}$, with $v_{\rm K}$ is
the Keplerian speed, and the functions
\begin{equation}
  \eta \equiv \frac{h^2}{2}\frac{d\log P}{d\log r}\,,\quad 
  \beta \equiv \sqrt{1+2\eta(r)}\,,
\end{equation}
and after neglecting the drag force in the momentum equations, the
exact background solution is
\begin{align}
  {\bf v}^{0}_{\textrm{g}} &=  \beta(r) {\bf v}_{\rm K} \label{eq:eq_disk_background1}\,,\\
  {\bf v}^{0}_{i} &= {\bf v}_{\rm K}\label{eq:eq_disk_background2}\,,
\end{align}
for $i=1,\dots,N$.

When the collision term between species is considered, as an
approximation we can assume that the velocity is slightly modified so it
can be written as the background solution
\eqref{eq:eq_disk_background1}-\eqref{eq:eq_disk_background2} plus a
small deviation, i.e., ${\bf v} = {\bf v}^{0} + \delta {\bf v}$.

Defining the function
\begin{equation}
  \xi \equiv \beta\left(\frac{1}{2} + \frac{d\log \beta}{d\log r}\right)\,,
\end{equation}
and neglecting terms which are second-order in the perturbations, the
steady-state axisymmetric Navier-Stokes equations lead to the
following set of algebraic equations for the perturbed velocities
\begin{align}
-2\beta \delta v_{{\rm g}\varphi} + \sum_{i=1}^N \frac{\epsilon_i}{{T_{\rm s}}_i} \left(\delta  v_{{\rm g}r} - \delta {v_i}_r\right) &= 0 \label{eq:eqdisk_gas1}\,, \\
\xi\delta  v_{{\rm g}r} + \sum_{i=1}^N \frac{\epsilon_i}{{T_{\rm s}}_i}\left(\delta  v_{{\rm g}\varphi} - \delta {v_i}_\varphi \right) &= \left(1-\beta\right) v_{\rm K} \sum_{i=1}^N \frac{\epsilon_i}{{T_{\rm s}}_i} \label{eq:eqdisk_gas2}\,,\\
-2 \delta {v_i}_\varphi + \frac{1}{{T_{\rm s}}_i} \left(\delta {v_i}_r-\delta  v_{{\rm g}r} \right) &= 0\,, \label{eq:eqdisk_dust1}\\
\frac{1}{2}\delta {v_i}_r + \frac{1}{{T_{\rm s}}_i}\left(\delta {v_i}_\varphi-\delta  v_{{\rm g}\varphi}\right) &= \frac{\left(\beta-1\right) v_{\rm K}}{{T_{\rm s}}_i}\,, \label{eq:eqdisk_dust2}
\end{align}
for $i=1, \dots, N$, with $T_{\rm s}$ the Stokes number (see \ref{eq:alpha_st}).
Eqs.\,\eqref{eq:eqdisk_gas1}-\eqref{eq:eqdisk_dust2} must be solved
coupled with the continuity equations
\begin{align}
  \partial_r\left(r\Sigma^0_{{\rm g}} \delta v_{{\rm g}r}\right) &= 0\,,   \label{eq:drift_continuty1}
  \\
  \partial_r\left(r\Sigma^0_{{i}} \delta v_{ir}\right) &= 0\,. \label{eq:drift_continuty2}
\end{align}

\medskip

From equations\,\eqref{eq:eqdisk_dust1} and\,\eqref{eq:eqdisk_dust2}
we obtain the dust velocities in terms of the gas velocity which, in
combination with \eqref{eq:eqdisk_gas1} and \eqref{eq:eqdisk_gas2},
allow us to find the gas velocity perturbations.
Defining
\begin{equation}
  \mathcal{S}_N \equiv \sum_{i=1}^N \frac{\epsilon_i }{1+{T_{\rm s}}_i^2}\,,\quad \mathcal{Q}_N \equiv \sum_{i=1}^N \frac{\epsilon_i {T_{\rm s}}_i}{1+{T_{\rm s}}_i^2}\,,
\end{equation}
the gas velocity perturbations read
\begin{align}
  {\delta v_{{\rm g}r}}(r) &= -2\beta\mathcal{Q}_N \Psi \left(\beta-1\right)v_{\rm K}\,, \label{eq:eq_disk_ur}\\
  {\delta v_{{\rm g}\varphi}}(r) &= -\left[ \left(\mathcal{S}_N + 2\xi\right) \mathcal{S}_N + \mathcal{Q}_N^2\right] \Psi \left(\beta-1\right)v_{\rm K} \label{eq:eq_disk_uphi}\,,
\end{align}
with $\Psi \equiv  \left[ \left(\mathcal{S}_N + \beta\right)\left(\mathcal{S}_N+ 2\xi \right) + \mathcal{Q}_N^2 \right]^{-1}$.

Finally, the expressions for the dust velocity perturbations are
\begin{align}
  {\delta v_i}_r &= \frac{2{T_{\rm s}}_i}{1 + {T_{\rm s}}_i^2}\left(\beta - 1\right)v_{\rm K} + \frac{ \delta v_{{\rm g}r} + 2{T_{\rm s}}_i \delta v_{{\rm g}\varphi}}{1 + {T_{\rm s}}_i^2} \label{eq:eq_disk_vr} \,,\\
  {\delta v_i}_\varphi &= \frac{1}{1 + {T_{\rm s}}_i^2}\left(\beta - 1\right)v_{\rm K} + \frac{2 \delta v_{{\rm g}\varphi}- {T_{\rm s}}_i \delta v_{{\rm g}r}}{2\left(1+{T_{\rm s}}_i^2\right)}\label{eq:eq_disk_vp}\,.
\end{align}

The velocities given by Eqs.\,\eqref{eq:eq_disk_ur} and
\eqref{eq:eq_disk_vp} are solution only if they satisfy the continuity
equations \eqref{eq:drift_continuty1} and \eqref{eq:drift_continuty2}.
For simplicity, in this work, we focus on the particular case of
non-flared disks (i.e., $h = h_0$, with $h_0$ being not necessarily
small), where $\beta = \beta_0$ and $\xi = \beta_0/2$. Thus, all the
velocity perturbations scale with the Keplerian speed, $v_{\rm K}$,
and the background surface-density profiles are power-laws with
exponent $d\log \Sigma / d\log r = -1/2$.

When considering only one dust species and $h\ll 1$, we can write
$\beta \simeq 1+\eta$ and \eqref{eq:eq_disk_ur}-\eqref{eq:eq_disk_vp}
are the solution found by \cite{Nakagawa1986}.  \cite{Dipierro2018}
found a similar solution for arbitrary number of species for a viscous
disk assuming a Keplerian background. This solution can be easily
improved following our approach.

\subsubsection{Numerical solution}
\vspace{-0.2cm}
We now use the steady-state solution found in the previous section to
test our implementation. For this test we initialize a large-scale 1D
disk using the first-order steady-state solutions, given by
Eqs.\,\eqref{eq:eq_disk_ur}-\eqref{eq:eq_disk_vp}. The computational
domain spans from $r=1$ to $r=100$, evenly spaced in a logarithmic
grid over 1024 points. We assume an isothermal equation of state.
We set boundary conditions equal to the steady-state
solution for all the species. The absence of perfect numerical
equilibrium at the beginning of the runs produces wave-patterns that
propagate in the disk and reach the boundaries of the mesh. To remove
these spurious waves from the active domain we use small damping zones
close to the boundaries \citep{Val-Borro2006}. These buffers extend
over a region such that $\Delta \Omega = 0.1$ for both the inner and
outer buffers \citep[see][]{Benitez-Llambay2016a}, and the damping
rate is set to one third of the local Keplerian frequency. We only
damp the density and radial velocity to the value given by the initial
condition.

\begin{figure*}
  \centering
  \includegraphics[scale=1.0]{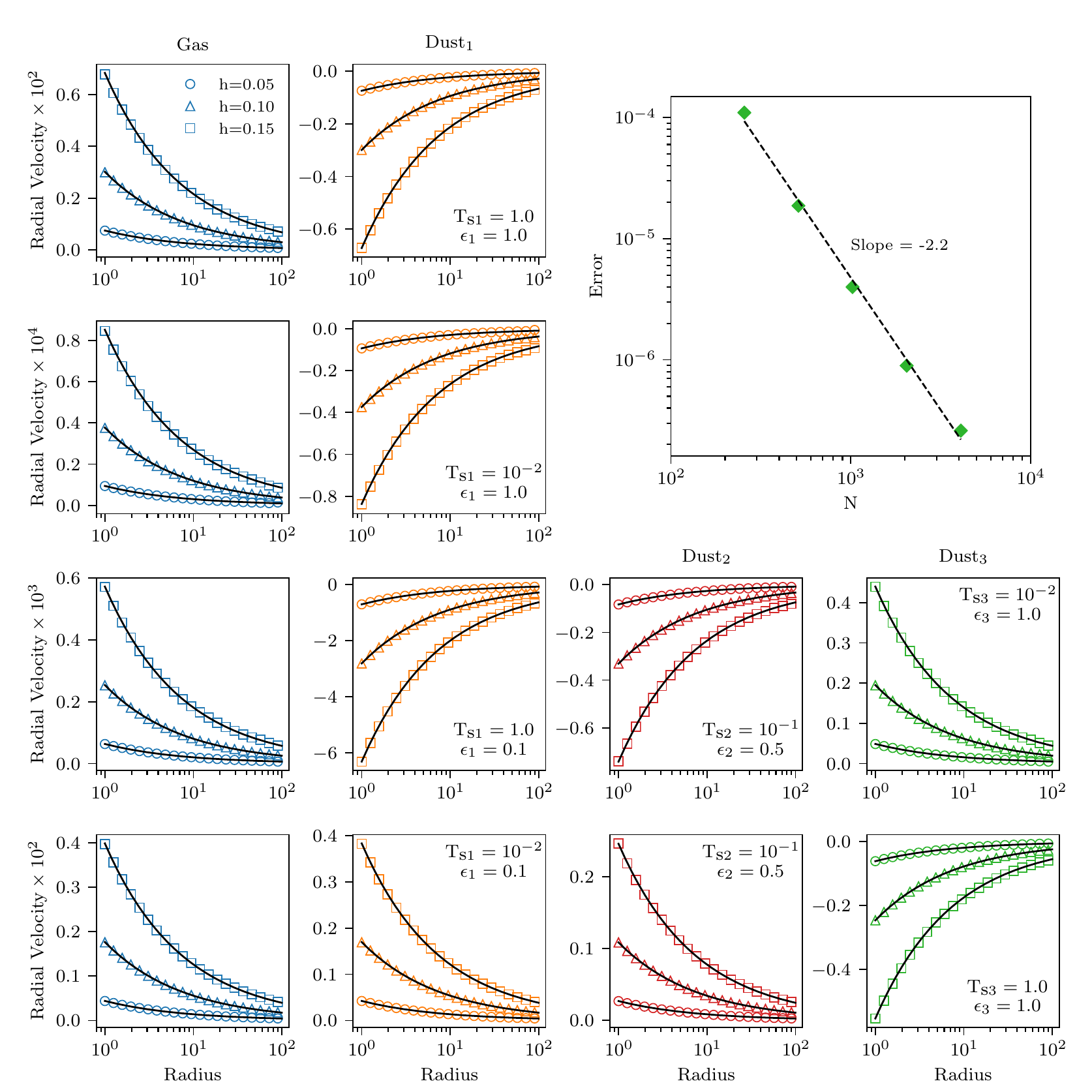}
  \caption{\label{fig:dustywave} Analytical (solid lines) and
    numerical (open colored symbols) solutions for the first-order
    dust radial drift test problem, described in Section
    \ref{sec:global_disk_solutions}. The analytical solutions are
    given by Eqs.\,\eqref{eq:eq_disk_ur} and \eqref{eq:eq_disk_vr}. In
    the smaller panels, we plot the radial velocity for all the cases
    studied. Different columns correspond to different species
    (labeled at the top of each of the uppermost panels) while
    different rows correspond to runs with different parameters. For
    each set of parameters, we run the same test with different
    aspect-ratios, $h$, and plot the resulting radial velocity with
    different symbols (see the legend in the leftmost upper
    panel). The parameters of each run are quoted inside of the small
    panels. In the large panel located at the right upper corner, we
    additionally plot the result from the convergence test described
    in Section \ref{subsec:global_disk_convergence}.  }
  \label{fig:eq_disk}
\end{figure*}

We consider two cases with two species and two cases with four
species, and vary the degree of coupling between gas and dust
species. To test our implementation in more challenging regimes, for
each configuration, we furthermore vary the aspect ratio, $h$, of the
disk, adopting the values $h\in\left[0.05,0.1,0.15\right]$. In order
to satisfy the hypothesis when deriving the analytical solution, we
work with non-flared disks. In all the cases, we numerically integrate
the 1D disks until the steady-state is reached. The initial surface
density of the gas component is not relevant for these tests.

In Fig.\,\ref{fig:eq_disk}, we plot the radial velocity for all the
cases studied. The results corresponding to different species and
different parameters are shown in each column and row,
respectively. From top to bottom, in the first two rows, we plot the
radial velocity for the two-fluid configurations and, in the last two
rows, the radial velocity for the test with four fluids. The
analytical solutions, given by Eqs.\,\eqref{eq:eq_disk_ur} and
\eqref{eq:eq_disk_vr}, are plotted with solid lines. The different
colors represent different species. Furthermore, depending on the
adopted aspect-ratio, we plot the data set using different
symbols. The parameters corresponding to each species are quoted
inside of the panels. We observe that the agreement between the
analytical and numerical solutions is excellent, and independent on
the parameters and the number of species.  The tests presented here
validate simultaneously the first-order steady-state disk-drift
solution and our numerical implementation.

We note an interesting result from the multiple fluid test. In the two
fluids cases, because of momentum conservation, it is impossible to
revert the sign of the radial velocity of the dust component but, its
magnitude depends on the dust-to-gas mass ratio as well as the degree
of coupling to the gas. However, this is no longer true in the more
general case of multiple species. In this case, very well coupled dust
can drift outward with the gas (see, for example, the fourth panel of
the third row and the second and third panels of the fourth row). We
finally comment that the same level of agreement was observed for the
azimuthal velocity, which is not surprising given that the two
directions are coupled.

\subsubsection{Convergence test}
\label{subsec:global_disk_convergence}

We additionally performed a convergence test with resolution. This
test consists in taking one particular case and measuring the error of
the numerical solution when changing the resolution.
For this particular case, we defined the error as:
\begin{equation}
\textrm{error}(\Delta r) = \frac{1}{N}\sum_{j=1}^N \left< \frac{v_{jr}^{\Delta r} - v_{jr}}{v_{jr}}\right>^2\,,
\end{equation}
with $N$ the total number of species and $v_{jr}^{\Delta r}$ the
solution obtained for different resolutions. We denote
the average over the cells with $\langle \rangle$.

For this test we take the case corresponding to the fourth row of
Fig.\,\ref{fig:eq_disk}. For this particular problem, we find that 256
cells are enough to obtain a converged solution.  We then use 256 as
starting number of grid points and go up to 4096, progressively
increasing by factors of 2.

We plot the result of this convergence test in the large panel of
Fig.\,\ref{fig:eq_disk}.  We successfully recovered a convergence rate
close to the expected order of the numerical method. For this test,
the time step was allowed to vary according to the CFL
condition. Thus, since the errors in space decrease rapidly, the
convergence rate is dominated by the first-order error in time.

\subsection{Streaming instability} \label{sec:SI_local}

The aerodynamic coupling between solids and gas in a differentially
rotating disk leads to the so-called streaming instability
\citep{Youdin2005}. Particular modes of this instability has been
extensively studied both in linear and non-linear regimes.  This is an
excellent problem to test our implementation both in the linear and in
the non-linear regimes. Due to the complexity of the problem and the
degree of coupling between all the equations and the directions, even
recovering linear solutions can be a stringent test.

In this paper, we extend previous studies in the linear regime to the
case of multiple dust species. In addition, for comparative purposes
with previous works, we show results in the non-linear regime
considering only one gas and one dust species.

\subsubsection{Preliminaries}

The growth rate of the streaming-instability can be obtained, in its
simplest form, by solving the 2.5D linearized axisymmetric
shearing-box equations for gas and one dust species. The fluid
equations are usually linearized around the steady-state drift
solution obtained by \cite{Nakagawa1986}.  However, in order to study
the instability for arbitrary number of dust species, generalized
background solutions are needed. These are similar to the approximated
solutions obtained for the a global disk (see Section
\ref{sec:global_disk_solutions}), but in this case are analytical and
exact. We derive and write them explicitly in Section \ref{sec:sb-eq}.

In the shearing-box approximation, a self-consistent aerodynamic drag
between gas and dust cannot be obtained. However, the instability can
still be studied in this formalism by adding an external constant
force mimicking the effect of a constant pressure gradient within the
box \citep[see e.g.][]{Bai2010}.

The equations leading to the streaming instability, when $N$ dust
species are considered, are
\begin{align} 
  \partial_t \rho_{\rm g} + \nabla \cdot\left( \rho_{\rm g} \mathbf{v}_{\rm g}\right) &= 0\,, \label{eq:streaming_gas_continuity} \\
  \partial_t \rho_{j} + \nabla \cdot\left( \rho_{j} \mathbf{v}_{j}\right) &= 0\,,
\end{align}
\begin{align}
  \partial_t \mathbf{v}_{\rm g} + \mathbf{v}_{\rm g} \cdot \nabla \mathbf{v}_{\rm g} =& - \frac{\nabla P}{\rho_{\rm g}} + \chi_0\Omega_0  \mathbf{e}_x + 2q\Omega_0^2 x \mathbf{e}_x \nonumber \\ 
  &-  2\mathbf{\Omega}_0 \times \mathbf{v}_{\rm g}- \Omega_0 \sum_{k=1}^{N}\frac{\epsilon_k\mathbf{\Delta}_k}{T_{{\rm s}k}}\,,   \\
  \partial_t \mathbf{v}_{j} + \mathbf{v}_{j} \cdot \nabla \mathbf{v}_{j} = & 2q\Omega_0^2 x \mathbf{e}_x - 2\mathbf{\Omega}_0\times \mathbf{v}_{j} + \Omega_0 \frac{\mathbf{\Delta}_j}{T_{{\rm s}j}}\,,\label{eq:streaming_dust_momenta}
\end{align}
for $j=1,\dotsc,N$, with $q$ the shear parameter. The term
$\chi_0\Omega_0$ is the constant radial acceleration that mimics the
pressure gradient within the box, with $\chi_0$ an arbitrary constant
speed. It is usually chosen to reproduce the drift speed of dust in
protoplanetary disks, i.e., $\chi_0 = 2h_0^2 v_{\rm K0}$, with $h_0 =
c_{\rm s 0}/v_{\rm K0}$ and $c_{\rm s0}$ the constant sound speed. The
unit vector along the radial direction is denoted as
$\mathbf{e}_x$. The pressure is related to the density as $P = c_{\rm
  s 0}^2\rho_{\rm g}$. The other terms depend on the dust-to-gas mass
ratio $\epsilon_i \equiv \rho_{i}/\rho_{\rm g}$, the Stokes number
$T_{{\rm s}i}$ and, the relative velocity vector between species
$\mathbf{\Delta}_i = \mathbf{v}_{\rm g}-\mathbf{v}_{i}$, where
$\mathbf{v}_{\rm g}$ and $\mathbf{v}_{i}$ are the gas and dust
velocity vectors respectively.

\begin{table*}
  \centering
  \caption{Eigenvalues, eigenvectors and parameters for the runs LinA, LinB and Lin3.}
  \label{tab:SI_eig}
  \begin{tabular}{cccc}
    \tablewidth{0pt}
    \hline\hline
    & LinA       & LinB                       & Lin3 \\
    \hline
    \decimals
    Parameters \\
    $K$                            & 30                          &   6                         &   50     \\
    $T_{{\rm s}1}$				   & 0.1						 &   0.1                       &   0.0425 \\
    $\epsilon_1$                   & 3.0						 &   0.2					   &   1.0    \\
    $T_{{\rm s}2}$				   & --						     &   --                        &   0.1    \\
    $\epsilon_2$                   & --						     &   --					       &   0.5    \\
    \hline
    Eigenvalue \\
    $\omega/\Omega_0$              & $-0.4190091323 + 0.3480181522i$ & $-0.0154862262  -  0.4998787515i$ & $-0.3027262829  + 0.3242790653i$      \\
    \hline
    Eigenvector \\
    $\delta \tilde{\rho}_{\rm g}$  & $+0.0000074637 + 0.0000070677i$ & $-0.0000337227 - 0.0003456248i$ & $+0.0000061052 + 0.0000080743i$    \\
    $\delta \tilde{v}_{{\rm g}x}$   & $-0.0563787907 + 0.0120535455i$ & $-0.0870451125 - 1.3851731095i$ & $-0.1587288108 + 0.0213251096i$    \\
    $\delta \tilde{v}_{{\rm g}y}$   & $+0.0445570113 + 0.0197224299i$ & $+1.3839936168 - 0.0937424679i$ & $+0.1327989476 + 0.0674232641i$    \\	    
    $\delta \tilde{v}_{{\rm g}z}$   & $+0.0563784989 - 0.0120536242i$ & $+0.0870497444 + 1.3852113520i$ & $+0.1587286212 - 0.0213252588i$    \\
    $\delta  \tilde{v}_{1x}$       & $-0.0466198076 + 0.0124333223i$ & $+0.2314730923 - 1.3715260043i$ & $-0.1461274403 + 0.0234873672i$    \\
    $\delta  \tilde{v}_{1y}$       & $+0.0435211557 + 0.0213517453i$ & $+1.3696536978 + 0.0196879160i$ & $+0.1325843682 + 0.0691301709i$    \\
    $\delta  \tilde{v}_{1z}$       & $+0.0546507401 - 0.0077776652i$ & $+0.0416164539 + 1.3844311928i$ & $+0.1571142133 - 0.0174328415i$    \\
    $\delta \tilde{\rho}_{2}$      & --                              &  --                             & $+0.1522281314 + 0.1836379253i$    \\
    $\delta  \tilde{v}_{2x}$       & --                              &  --                             & $-0.1335593453 + 0.0025396632i$    \\
    $\delta  \tilde{v}_{2y}$       & --                              &  --                             & $+0.1092222067 + 0.0952973332i$    \\
    $\delta  \tilde{v}_{2z}$       & --                              &  --                             & $+0.1485545469 + 0.0200753935i$    \\
    \hline
  \end{tabular}
  \smallskip\newline \tablecomments{The dimensionless velocity amplitudes
    and wavenumber are defined as $\delta \tilde{v} = \delta v/(h_0^2v_{{\rm K}0})$ and 
    $K = kh_0^2v_{\rm K 0}/\Omega_{0}$, respectively (see Appendix \ref{ap:SI}). The dust-density perturbation $\delta \tilde{\rho}_{1} = 1$ for all the runs.}
  \medskip
\end{table*}

\subsubsection{Steady-state solution}
\label{sec:sb-eq}

As discussed above, when setting a constant background density for the
gas and all the dust species, an exact steady-state solution can be
found. This solution is the generalization of that obtained by
\cite{Nakagawa1986}, and the procedure to find it is similar to that
followed when finding the solution for the perturbations in Section
\ref{sec:global_disk_solutions}. Defining
\begin{align}
\mathcal{A}_N &=  \tilde{\kappa}^2\sum_{i=1}^{N} \frac{\epsilon_iT_{{\rm s}i}}{1+\tilde{\kappa}^2 T_{{\rm s}i}^2} \, \,,\\
\mathcal{B}_N &= 1 + \sum_{i=1}^{N}  \frac{\epsilon_i}{1+\tilde{\kappa}^2 T_{{\rm s}i}^2},
\label{eq:equi_gas_1}
\end{align}
the steady-state solution of Eqs.\,\eqref{eq:streaming_gas_continuity}-\eqref{eq:streaming_dust_momenta}
is
\begin{align}
v^{\textrm{0}}_{{\rm g}x} &= \mathcal{A}_N \chi_0 \psi \,, \label{eq:equi_gas} \\
v^{\textrm{0}}_{{\rm g}y} &=  -q\Omega_0x - \frac{\tilde{\kappa}^2}{2} \mathcal{B}_N \chi_0 \psi\,,
\end{align}
with $\psi = \left(\mathcal{A}_N^2 +
\tilde{\kappa}^2\mathcal{B}_N^2\right)^{-1}$ and $\tilde{\kappa}^2 =
\kappa^2 \Omega_0^{-2}$, where $\kappa^2 =
2\left(2-q\right)\Omega_0^2$, is the square of the epicyclic
frequency.

For the $i$-th dust species, its velocity can be written in terms of
the velocity of the gas as
\begin{align}
v^0_{i x } &= \frac{v^0_{{\rm g} x} + 2T_{{\rm s}i}\left(v^0_{{\rm g}y} + q\Omega_0x\right)}{1+\tilde{\kappa}^2 T_{{\rm s}i}^2}\,, \\
v^0_{i y } &=  -q\Omega_0x + \frac{\left(v^0_{{\rm g}y} + q\Omega_0x\right) - (2-q) T_{{\rm s}i}v^0_{{\rm g} x}}{1+\tilde{\kappa}^2 T_{{\rm s}i}^2}\,.
\label{eq:equi_dust}
\end{align}
The vertical velocities are $v^0_{{\rm g}z} =
v^0_{iz} = 0$ and the densities are constant for all the
species. We note that, for the case $q=3/2$ (i.e., Keplerian shear),
Eqs.\,\eqref{eq:equi_gas}-\eqref{eq:equi_dust} are, as expected,
equivalent to the expansion of
Eqs.\,\eqref{eq:eq_disk_ur}-\eqref{eq:eq_disk_vp} for $h_0 \ll 1$.

\subsubsection{Linear regime - eigenvalues and eigenvectors}

Assuming solutions of the form $\delta \rho = \rho^0 + \delta \rho\,,
{\bf v} = {\bf v}^0 + \delta {\bf v}\,,$ and after neglecting
quadratic terms in the perturbations,
Eqs.\,\eqref{eq:streaming_gas_continuity}-\eqref{eq:streaming_dust_momenta}
become a set of linear partial differential equations for the
perturbations $\delta$. Without loss of generality, we assume
perturbations of the form $\delta f(x,z,t) = \delta \hat{f} e^{i
  \left(k_x x + k_z z\right) - \omega t }$, from which
\eqref{eq:streaming_gas_continuity}-\eqref{eq:streaming_dust_momenta}
transform into a set of linear algebraic equations of the form
\begin{equation}
  {\bf A} {\bf u} = \tilde{\omega} {\bf u}\,,
  \label{eq:eigen-si}
\end{equation}
with ${\bf u}$ the column vector whose elements are the perturbation
amplitudes and $\tilde{\omega} = \omega/\Omega_0$ the normalized
eigenvalues. The problem then reduces to finding the normalized
eigenvalues and eigenvectors of the $N\times N$ matrix ${\bf A}$. We
write the explicit expression of this linear system in Appendix
\ref{ap:SI}.

A general expression for the dispersion relation and its eigenvectors
can be easily obtained and written in closed form (similarly to what
it was done in Section \ref{sec:sound-wave}). However, due to the
complexity of these expressions, we avoid writing them here. Instead,
when solving the eigenvalue problem, we simply write the matrix ${\bf
  A}$ and find its eigenvalues and eigenvectors numerically.

In Table \ref{tab:SI_eig} we present the parameters, the eigenvalues
and the eigenvectors for the three different cases studied in this
paper, called LinA, LinB, and Lin3. The first two cases correspond to
one gas and one dust species, and have already been studied
\citep[e.g.][]{Youdin2007,Balsara2009,Bai2010}. The third one contains
one gas and two dust species.

\begin{figure*}[htb!]
  \centering
  \includegraphics[scale=0.98]{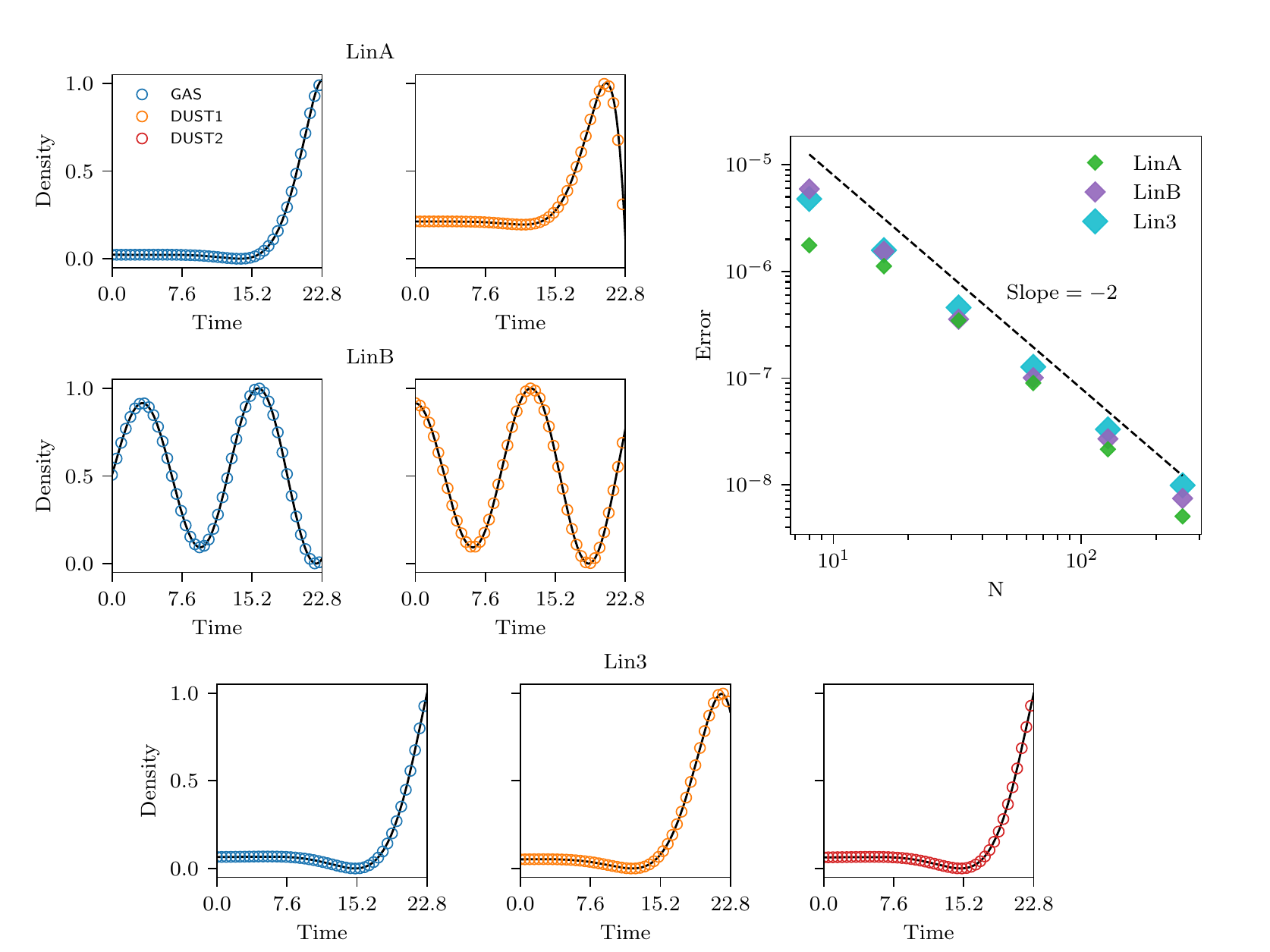}
  \caption{Analytical (solid lines) and numerical (open circles)
    solutions of the linear streaming instability, described in
    Section \ref{subsec:si-linear}, for the runs LinA (top), LinB
    (center) and Lin3 (bottom), obtained with $128^2$ grid
    points. From left to right, we plot the gas and dust densities. We
    additionally plot the result of the convergence test, described in
    Section \ref{subsec:si-linear_convergence}. The agreement between
    the analytical and numerical solutions is excellent. The slope
    recovered from the convergence test is consistent with the
    expected convergence rate for all the cases, showing small
    deviations for very low resolutions.}
  \label{fig:SI}
\end{figure*}

We report a small difference with respect to the eigenvalues obtained
by \cite{Youdin2007}. We tracked down the difference to two terms in
the linearized equations (see Appendix \ref{ap:SI}):
\begin{equation}
- \delta \tilde{\rho}_{\rm g} \sum_{k=1}^{N}\frac{\epsilon^0_k {\Delta}^0_{kx}}{T_{{\rm s}k}}\,, \quad \textrm{and} \quad - \delta \tilde{\rho}_{\rm g} \sum_{k=1}^{N}\frac{\epsilon^0_k {\Delta}^0_{ky}}{T_{{\rm s}k}}\,.
\end{equation}
Neglecting these terms modifies the fourth digit of the eigenvalues,
and allows us to recover the values reported by \cite{Youdin2007}.

\subsubsection{Linear regime - numerical solution}
\label{subsec:si-linear}

\begin{table*}[htb!]
	\centering
	\caption{Measured growth rates for different number of cells for the runs LinA, LinB and Lin3.}
	\label{tab:SI_growthrate}
	\begin{tabular}{cccc}
		\hline\hline
		N & LinA       & LinB                       & Lin3 \\
		\hline
		8      & $-0.325$    $\pm 3.3 \times 10^{-2}$ & $0.0301$      $\pm 1.2 \times 10^{-3}$ & -0.222    $\pm 8.5 \times 10^{-2}$ \\
		16     & $-0.3961$   $\pm 1.7 \times 10^{-3}$ & $-0.00821$    $\pm 2.9 \times 10^{-4}$ & -0.271    $\pm 4.9 \times 10^{-2}$ \\
		32     & $-0.41311$  $\pm 5.2 \times 10^{-4}$ & $-0.014468$   $\pm 7.3 \times 10^{-5}$ & -0.291    $\pm 1.3 \times 10^{-2}$ \\
		64     & $-0.41762$  $\pm 1.5 \times 10^{-4}$ & $-0.015349$   $\pm 2.2 \times 10^{-5}$ & -0.3000   $\pm 2.4 \times 10^{-3}$ \\
		128    & $-0.418583$ $\pm 8.0 \times 10^{-5}$ & $-0.0154688$  $\pm 6.5 \times 10^{-6}$ & -0.30248  $\pm 1.4 \times 10^{-4}$ \\
		256    & $-0.418900$ $\pm 5.4 \times 10^{-5}$ & $-0.0154839$  $\pm 2.0 \times 10^{-6}$ & -0.302672 $\pm 5.1 \times 10^{-5}$ \\
		\hline
	\end{tabular}
	\smallskip \tablecomments{The values correspond to the average
          of the growth rates obtained by fitting the time evolution
          of each component of the eigenvector. The errors correspond
          to the standard deviation of this average.}
	\medskip
\end{table*}

To test our implementation, we solve the fully non-linear set
of equations in the shearing-box approximation. We then compare the
results with those obtained from the linear solution described in the
previous section.

To numerically recover the solutions we set $h_0 = 0.05$ and ${v_{\rm
    K}}_0 = 1$, and add the constant external force $\chi_0$ to the
gas component along the $x$ direction.  The shear parameter $q$ is set
to $3/2$.  We only consider wavenumbers $k_x = k_z = k$, so we employ
a square axisymmetric shearing-box with $x,y\in\left[-L/2,L/2\right]$
and $L = 2\pi/k$. The grid is evenly spaced over 256 cells in each
direction. We set periodic and shear-periodic boundary conditions in
the $z$ and $x$ directions, respectively. The initial condition is
given by the steady-state background solution
\eqref{eq:equi_gas}-\eqref{eq:equi_dust}, and we set the background
densities to $\rho^0_{{i}} = \epsilon_i \rho^0_{{\rm g}}$, with
$\rho^0_{\rm g} = 1.0$.

Because of truncation errors, the numerical equilibrium does not
match, to machine precision, that given by
Eqs. \eqref{eq:equi_gas}-\eqref{eq:equi_dust}, but it is very
close. However, after initializing each run, the system quickly
relaxes toward an exact numerical equilibrium. Thus, to improve our
measurements, we wait for a time $t_0=1.2\Omega_0^{-1}$ until the
numerical equilibrium is obtained, and then excite the unstable
mode. We note that to speed-up the calculations, the relaxation step
can be done in a 1D grid. We fix the CFL factor to $0.3$ for all the runs.

The linear mode is excited by adding to the steady-state background
$f$ the small perturbation $\delta f$, defined as:
\begin{align}
  \delta f = & A \left[\operatorname{Re}\left(\delta \hat{f}\right) \cos(k_x x + k_zz) \right. \nonumber \\
    & \left. - \operatorname{Im}\left(\delta \hat{f}\right) \sin(k_x x + k_zz)\right],
\end{align}
where $\delta \hat{f}$ is the complex amplitude of the corresponding
component of the unstable eigenvector (see Table \ref{tab:SI_eig}) and
$A$ is a small amplitude that ensures linearity. Its value is set to
$A = 10^{-5}$.

\begin{figure*}[htb!]
  \centering
  \includegraphics[scale=1.0]{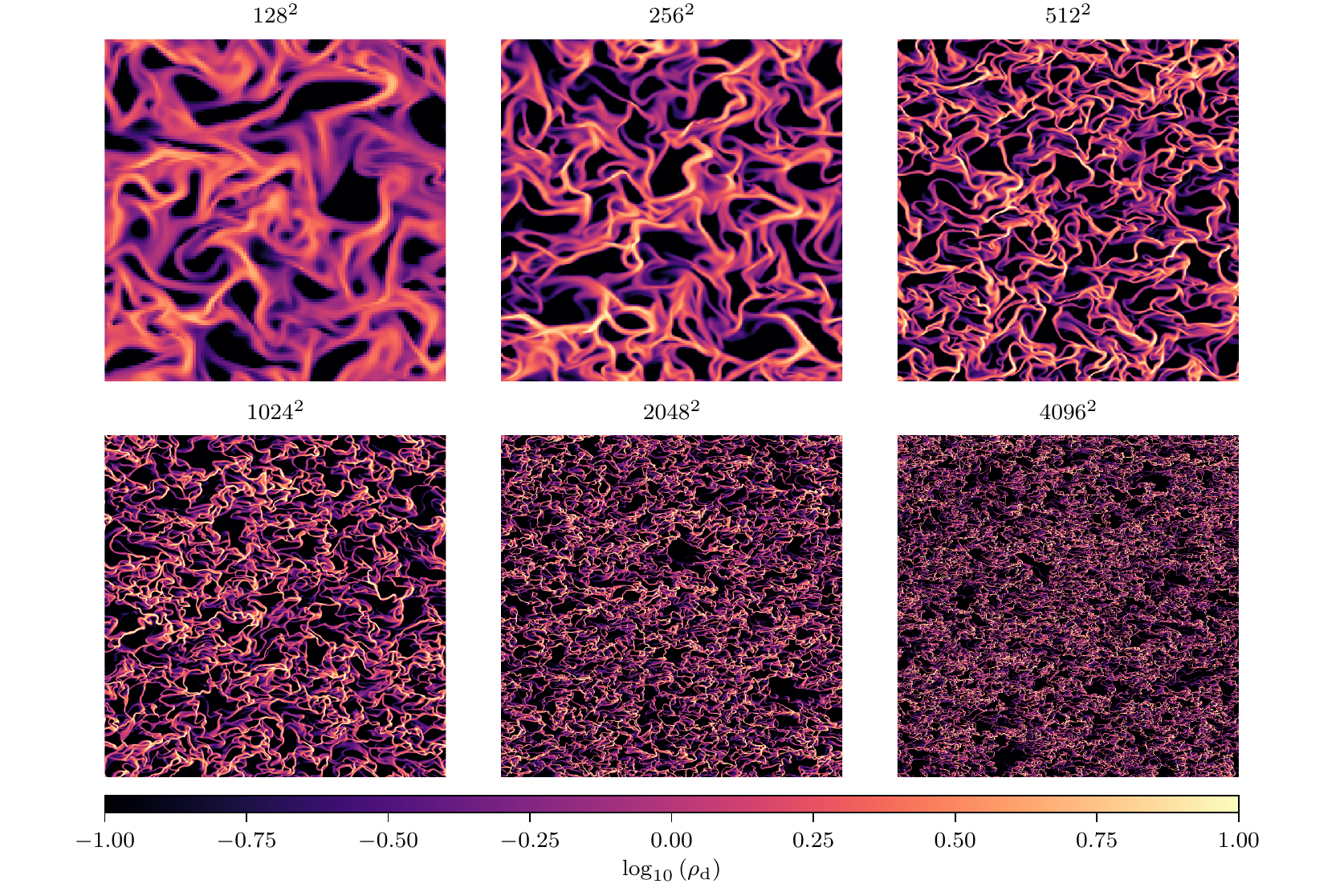}
  \caption{ Dust density maps for the test AB. Each panel is labeled
    by the total number of cells of the box. This mode is dominated by
    over-dense filaments and voids. The larger the resolution, the
    smaller and denser the filaments become. Convergence with
    resolution is far from being observed for the resolutions
    studied. The panels corresponding to $256^2$ and $1024^2$ cells
    can be compared with Fig.\,5 of \cite{Bai2010}, where a good
    qualitative agreement is observed.}\label{fig:SIAB}
\end{figure*}

In Fig.\,\ref{fig:SI}, we plot the time evolution of the normalized
density perturbations for each of the three different cases, measured
from time $t_0$ at the location $x = z = -L/2$ (this is an arbitrary
choice). The normalization is such that the density perturbation is
between zero and one for the time interval considered.  In each panel
we plot, with open circles, the values obtained numerically with our
implementation, while the solid lines are the analytical ones. The
color represents different species, blue being the gas and, orange and
red, the dust species.  The first two rows of Fig.\,\ref{fig:SI}
correspond to the tests LinA and LinB, respectively. The third one
corresponds to the three species Lin3 test. In all of the runs, the
agreement between the analytical and numerical solutions is
excellent. We additionally comment that the same level of agreement is
observed for the velocities of the gas and the dust species.

In Table \ref{tab:SI_growthrate}, we present the result of the
measured growth rates for the tests LinA, LinB and Lin3, for different
resolutions. The growth rate for each mode was obtained first by
fitting each component of the eigenvector and then averaging the
results of the fits. For the tests LinA and Lin3, the instability can
be recovered with 8 cells. However, for the mode linB, at least 16
cells are required to obtain an unstable behavior. The errors
correspond to the standard deviation of the average.

This test allows us to confidently conclude that our implementation is
correct and very robust. In the next section, we additionally study
the convergence rate for these test problems.

\subsubsection{Linear regime - Convergence test}
\label{subsec:si-linear_convergence}

To test the convergence rate of these test problems, we perform a
series of runs decreasing the resolutions by factors of two, starting
with $256^2$ cells down to $8^2$ cells.

We measure the convergence rate for the three configurations described
in the previous section by computing the error, defined as
\begin{equation}
  \textrm{error} = \left( \sum_{i=1}^{m} \left< \left(\delta f_i^{\Delta}(t) - \delta f_{i}(t) \right)^2\right> \right)^{1/2}\,,
\end{equation}
where $m$ is the number of components of the eigenvector, $\delta
f^{\Delta}$ the numerical solution, $\delta f_i$ is the analytical
one, and $\langle \, \rangle$ the time average between $t = t_0$ and
$t = 7\Omega_0^{-1}$.

The rightmost large panel of Fig.\,\ref{fig:SI}, shows the result of
the convergence test for the three different cases. We additionally
plot (dashed line) the expected second-order accuracy slope. The
lowest resolution cases slightly depart from it. However, an excellent
convergence rate is observed for $N>32^2$ grid points. The convergence
properties for all the modes analyzed demonstrate the validity of our
implementation. It is remarkable that, even with low resolution, our
implementation is able to recover the linear growth rate with an
acceptable level of accuracy.

We report that we have observed the mode LinB to be prone to develop
noise at cell level which, eventually, contaminates the computational
domain. By disabling the drag term, we have concluded that this noise
is something entirely related to the gas component. This issue was
significantly reduced by enabling a predictor using a half
transport-step before the source step, allowing us to recover
excellent second-order accurate linear solutions (see
\ref{subsec:implementation}).

\begin{figure*}[htb!]
  \centering
  \includegraphics[scale=1.0]{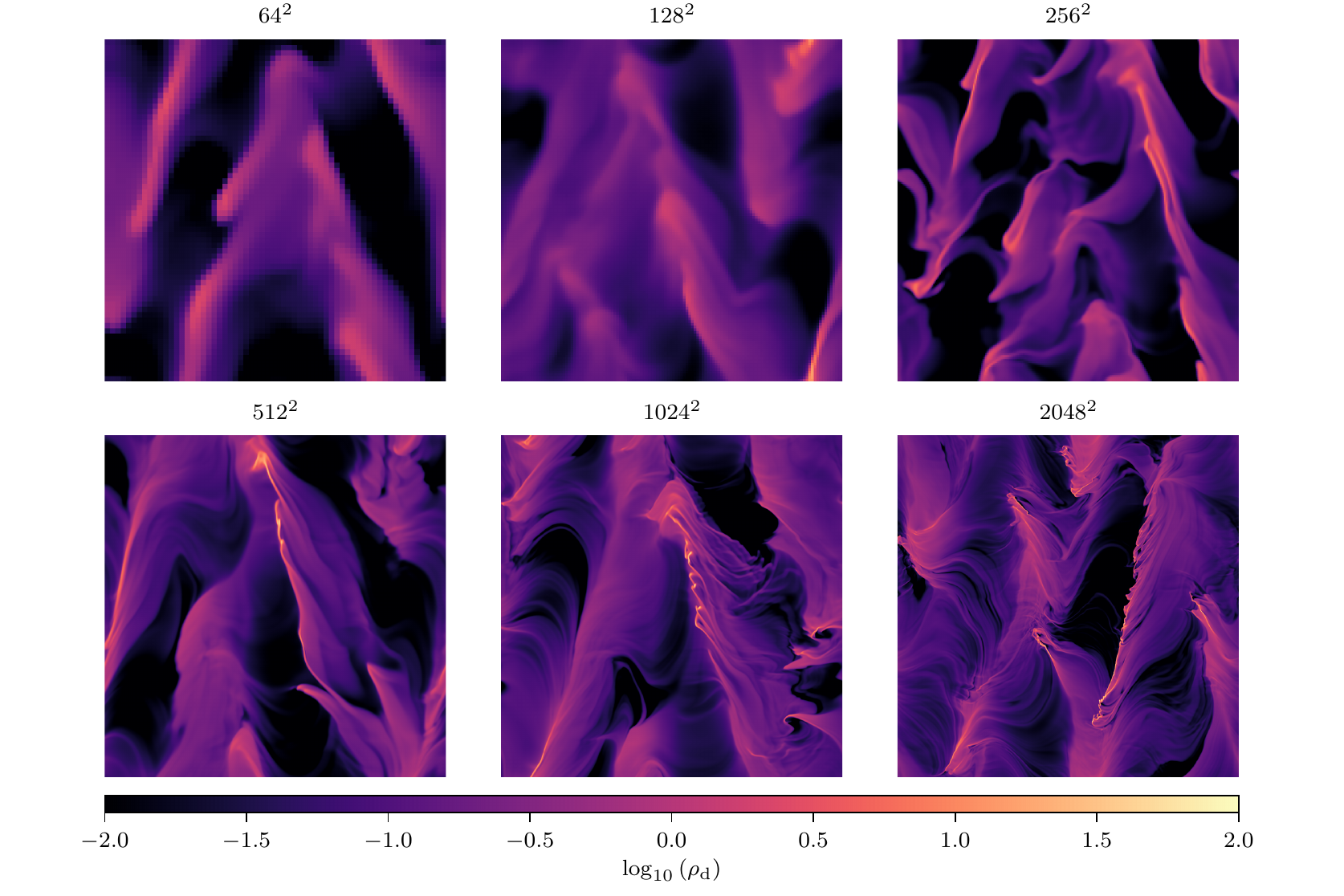}
  \caption{Dust density maps for the test BA. Each panel is
    labeled by the total number of cells of the box. While the
    number of details increases with the number of cells,
    convergence with resolution is observed for number of cells
    $>512^2$. The panels corresponding to $256^2$ and $1024^2$
    cells can be compared with Fig.\,5 of \cite{Bai2010}, where
    a good qualitative agreement is observed.}\label{fig:SIBA}
\end{figure*}

\subsubsection{Non-linear regime}

\label{sec:SI_nonlinear}

To study the non-linear regime of the streaming instability,
we consider the runs AB and BA described by \cite{JY07} and
\cite{Bai2010}. We focus our attention on the convergence with
resolution, the cumulative dust density distribution and the time
evolution of the maximum density. These tests give us, in particular,
the opportunity to assess whether the Eulerian approach for the dust
species is able to reproduce similar features as those obtained by
\cite{Bai2010} using Lagrangian particles.

For each test, we set a square shearing-box of size $L = l h_0 H_0$,
with $h_0 = 0.05$, $H_0 = h_0R_0$, and the fiducial radius $R_0=1$.
The shear parameter $q$ is, as above, $3/2$ (i.e., Keplerian
rotation).  For the test AB (BA), we set the dust-to-gas mass ratio
$\epsilon_1=1$ ($\epsilon_1 = 0.2$), the Stokes number $T_{{\rm
    s}1}=0.1$ ($T_{{\rm s}1}=1.0$), and the parameter $l = 2$ ($l =
40$).  The total integration time is set to $40\Omega_0^{-1}$ ($800
\Omega_0^{-1}$), which allows the saturated turbulent state to be
reached \citep{Bai2010}. We seed the instability with white noise in
the three velocity components of the two species, with an amplitude $A
= 10^{-2} h_0 v_{{\rm K }0}$.

To test convergence with resolution, for a fixed box size, we vary the
number of grid cells by a factor of four. For the test BA, we set the
nominal box with $64^2$ cells -- a resolution of roughly $32/H_0$ --
and obtain results when varying the number of cells up to $2048^2$ --
a resolution of $1024/H_0$. For the test AB, since a box with $64^2$
cells does not allow the instability to growth, we start with $128^2$
cells -- a resolution of $1280/H_0$ -- and increase it up to $4096^2$
cells -- a resolution of $40960/H_0$. We note that, when using $64^2$
cells for the run AB, \cite{Bai2010} were able to recover an unstable
evolution, which is probably due to the higher order of the Athena
code. For the run AB, and the lowest resolution ($128^2$ grid cells),
we report a saturation time $\simeq 12\Omega_0^{-1}$, a value higly
dependent on resolution. On the other hand, for the case BA, and the
lowest resolution ($64^2$ grid cells), it saturates after $\simeq
150\Omega_0^{-1}$. This value is not very dependent on resolution.

In Figs.\,\ref{fig:SIAB} and \ref{fig:SIBA}, we show snapshots of the
dust density when the instability is saturated, at times
$20\Omega_0^{-1}$ and $400 \Omega_0^{-1}$, for the runs AB and BA,
respectively.
\begin{figure*}[htb!]
  \centering
  \includegraphics[]{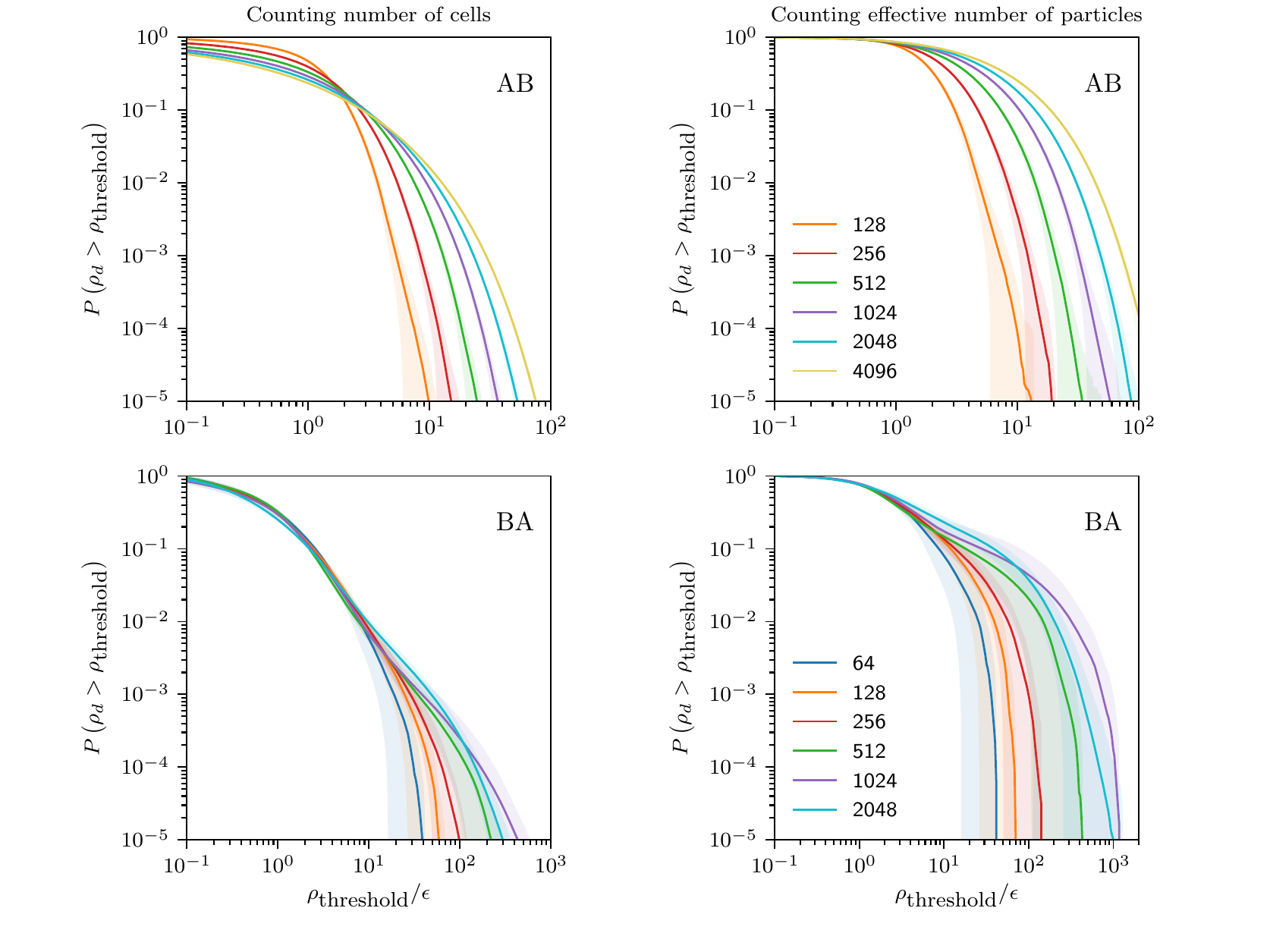}
  \caption{Cumulative dust density distributions for the models AB
    (top) and BA (bottom). Solid lines correspond to the time-averaged
    cumulatives. Shaded regions correspond to the standard
    deviation. The different colors represent each of the cases shown
    in Fig.\,\ref{fig:SIAB} and \ref{fig:SIBA}.  The left and right
    panels show the distributions obtained by counting cells and
    density, respectively. The distributions are normalized such that
    they integrate to one (left panels) or the probability of the
    lowest density-threshold is equal to one (right panels).  The
    upper panel shows that, for the mode AB, the maximum density
    increases linearly with the resolution, a clear evidence of lack
    of convergence. Contrary to this case, the bottom panel shows
    that, for a number of cells $>512^2$, the mode BA converges for
    all the density values. These results are independent of the
    statistical method used to compute the distributions. The right
    panels can be directly compared with Fig. 6 of \cite{Bai2010}.}
    \label{fig:particle_density}
\end{figure*}
Fig.\,\ref{fig:SIAB} shows that, for the test AB, smaller and denser
structures develop when the resolution increases, where no sign of
convergence with resolution is observed. This effect, while still
present, is not so strong for the low resolution runs in the test BA
(Fig.\,\ref{fig:SIBA}). Naively, this can be understood by analyzing
the dispersion relation of the instability \citep[see
  e.g.][]{Youdin2007}. In the absence of any dissipative process, such
as viscosity or diffusion, the smaller scales ($k_z \rightarrow
\infty$) grow at a rate given by the maximum growth rate. Thus,
density concentrations are prone to grow in very localized regions, a
trend that can be clearly recognized in Fig.\,\ref{fig:SIAB} for the
case AB. We refer to this as a naive explanation because it is not
clear that the same occurs for the case BA, even when considering that
the dispersion relation is not very different from that obtained for
the case AB. Further studies are necessary to understand the real
source of the discrepancy in the convergence properties between these
two cases.

The panels that correspond to $256^2$ and $1024^2$ cells can be
compared with those presented in Fig.\,5 of \cite{Bai2010}. The level
of qualitative agreement between the dust density obtained using a
particle approach \citep{Bai2010} and our fluid approach is
remarkable. We note that, in the non-linear turbulent regime, the
instability could, in principle, be dominated by crossing
trajectories, thus invalidating our approach. However, the overall
agreement obtained from this qualitative comparison suggests that the
dynamics of the instability, in the non-linear regime, could be
treated using a fluid approach.

To better quantify the convergence properties for both the AB and BA
tests, following \cite{Youdin2007} and \cite{Bai2010}, we study the
cumulative dust-density distribution. We calculate it by following two
different procedures, one by the counting number of cells with density
above some threshold value, $\rho_{\textrm{threshold}}$, and another
one by adding up the density of cells with density above
$\rho_{\textrm{threshold}}$. The latter is similar to counting (the
effective) number of particles, as done by \cite{Youdin2007} and
\cite{Bai2010}.  We split the dust density in 300 logarithmic bins,
between $\log_{10} (\rho_{a}\epsilon)$ and $\log_{10}
(\rho_{b}\epsilon)$, where $\rho_{b}=10^{2}$ for the case AB, while
$\rho_{b}=2\times10^{3}$ for BA, and $\rho_{a} = 10^{-1}$ in both
cases. To obtain a representative cumulative function of the saturated
regime, we compute it for different times, between $t=30\Omega_0^{-1}$
and $t=600\Omega_0^{-1}$ for the cases AB and BA, respectively, until
the final integration time, and we finally average them. We also
compute the standard deviation, which provides valuable information
about the fluctuations of the density in the saturated phase.

\begin{figure*}[htb!]
  \centering
  \includegraphics[]{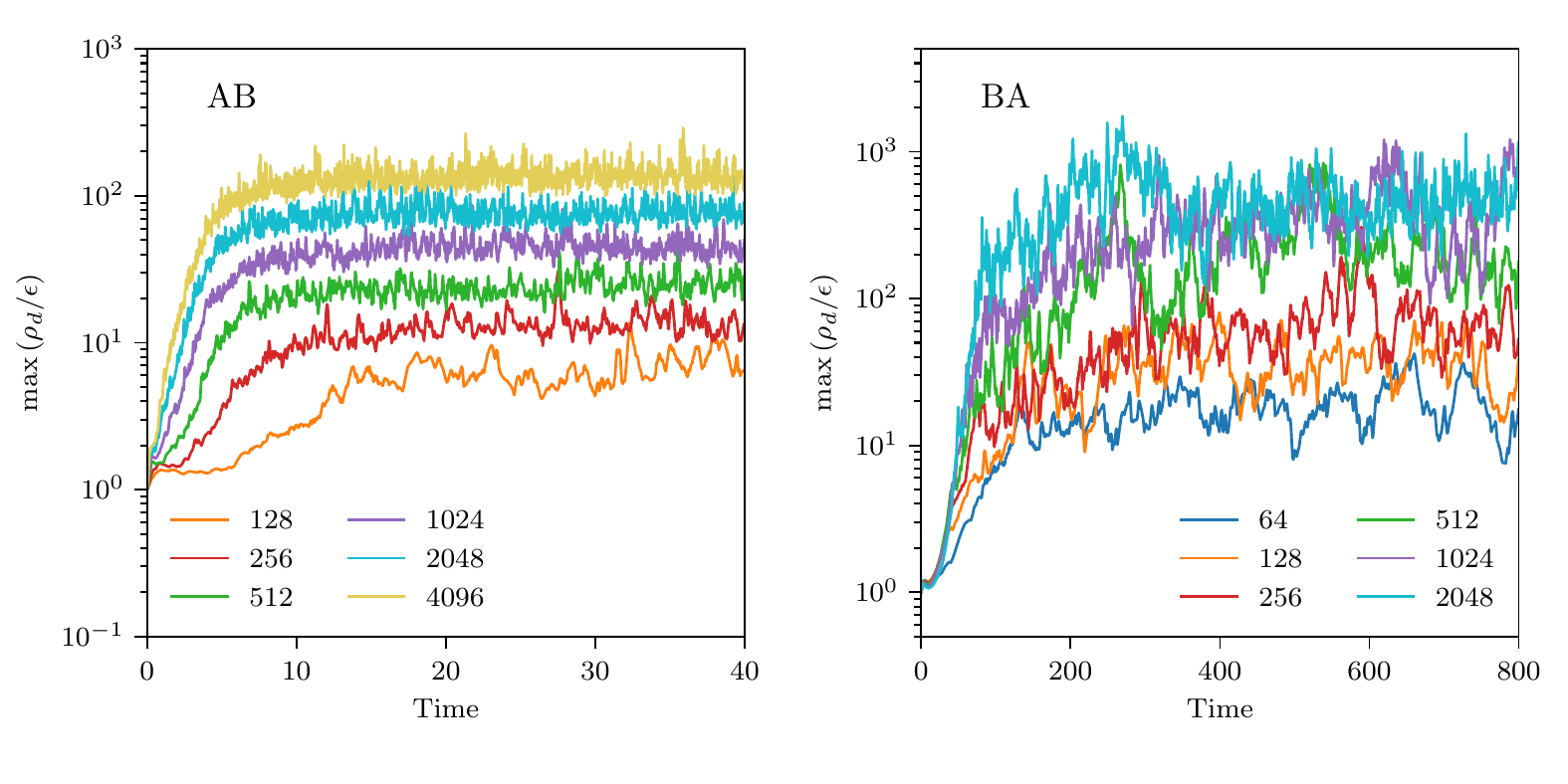}
  \caption{Maximum dust density over time for the modes AB (left
    panel) and BA (right panel). As shown in
    Fig.\,\ref{fig:particle_density}, the maximum density for the case
    AB increases linearly with resolution. The left panel also shows
    that the time for saturation is directly correlated with the
    resolution. Contrary to this case, the mode BA presents much
    better convergence properties with resolution. The right panel
    shows that, for low resolution, the initial growth rate directly
    correlates with resolution. However, for number of cells $>512^2$,
    convergence in the growth rate is observed. The same degree of
    convergence is observed for the maximum dust density over the time
    interval considered, where both the maximum and fluctuations are
    comparable.}
  \label{fig:SImax}
\end{figure*}

In Fig.\,\ref{fig:particle_density} we plot the time averaged
cumulative distributions for the dust density, corresponding to the
cases AB (upper panels) and BA (lower panels). In each panel, and with
different colors, we plot the cumulative function corresponding to the
data shown in the panels of Fig.\,\ref{fig:SIAB} and \ref{fig:SIBA}.
Shaded regions show the standard deviation.  The left and right panels
show the results obtained by counting the number of cells and by
summing the density of the cells, respectively. In the left panels,
the distributions are normalized such that they integrate to one. For
comparative purposes with \cite{Bai2010}, the curves in the right
panels are normalized such that the probability of the minimum density
bin is one.

For both cases, AB and BA, the dispersion is very small and does not
depend on the method used to calculate the cumulative distribution. In
particular, for the run BA, a strong degree of convergence, down to
probabilities of the order $P(\rho_{\rm d}>\rho_{\textrm{threshold}})
\sim 10^{-3}$ is observed for all the resolutions.  We report that,
for $P=10^{-5}$, the obtained probabilities correspond to values of
$\rho_{\textrm{threshold}}/\epsilon$ that are roughly one order of
magnitude below the values presented by \cite{Bai2010} when counting
number of cells. However, counting (the effective) number of
particles, by adding up densities, removes this discrepancy. Contrary
to what is observed for the case AB, the mode BA seems converged for a
number of cells larger than $512^2$.

The method used to calculate the cumulative distributions does not
modify the degree of convergence found for each run, AB and
BA. However, the shape of the distributions is method-dependent.  When
counting cells, lower densities contribute more significantly in
shaping the cumulative distribution, while when counting (the
effective) number of particles, denser regions contribute more. While
the maximum-density values differ from those obtained by
\cite{Bai2010} for the run AB, the overall shape of the distributions
agrees better when adding up densities.

The run AB shows a direct correlation between the saturation time
scale and the resolution, i.e., the higher the resolution, the faster
the instability saturates.  Furthermore, we find the maximum density
to be proportional to the number of cells, a clear evidence of lack of
convergence.  The previous analysis is supported by
Fig.\,\ref{fig:SImax}, where we plot the maximum dust density as a
function of time for each case and resolution.  As described above, we
show again that the run BA is much better behaved in terms of
convergence. While the maximum density also increases with resolution
for the low resolution cases, it converges when using more than
$512^2$ cells.

The differences found when comparing our results with those obtained
using Lagrangian particles, in particular the lack of convergence for
the run AB, warrant a detailed comparison between these two
approaches.

\bigskip

\bigskip

\bigskip

\section{Summary and perspectives} 
\label{sec:sec4}

In this paper, we have presented a reliable numerical method developed
to solve the momentum transfer between multiple species. We have
focused on systems composed of gas and several dust species in the
linear drag regime, which has a broad spectrum of applicability in
studies of protoplanetary disks. Nevertheless, we have also shown in
Appendix\,\ref{ap:NL}, by means of a simple example, how this method
could be extended to the non-linear drag regime.

The core of our implementation is the correct treatment of the
coupling term between different species. This is solved using a
first-order fully implicit method and connected to the fluid solvers
via the operator splitting approximation. After analyzing different
test problems of varying complexity, we have shown that the code
conserves its second-order accuracy in space.

The implicit scheme was designed to conserve momentum to machine
precision, a quantity that must be conserved during collisions between
pairs of species. This property is fundamental to correctly describe
the physical evolution of a system of multiple species and makes the
solver extremely robust. In addition, we have shown that the implicit
scheme is asymptotically and unconditionally stable, with the correct
asymptotic limits. Furthermore, this property is independent of the
number of species.

Since the value of the Stokes number is not directly involved when
studying the properties of the implicit scheme, we stress that the
algorithm works well independently of the Stokes number, as it was
confirmed by our test suite.

Before using this implementation to study a particular physical
problem, it is necessary to assess for which Stokes numbers the fluid
approximation is expected to produce a good description of the
dynamics of the problem. Generally, if the dynamics of the problem is
not dominated by crossing trajectories -- which are transformed into
shocks by the fluid approach -- the fluid approach should remain as a
good approximation.

A qualitative comparison between \cite{Bai2010} and this work, has
shown that the agreement between the particles and the pressureless
fluid approximation for dust is very good, while we were not able to
find convergence for the run AB. Further studies are still needed in
order to assess the level of agreement between both approximations.

In this paper we did not take into account the thermal evolution of
the species and the possibility of mass transfer between them. These
two ingredients are very important in order to study the
self-consistent dust evolution in protoplanetary disks. In particular,
to properly account for the thermodynamics of the system, models for
frictional heating, heat conduction, and radiation should be
considered for the gas and dust species.

This implementation has already been successfully used to study
problems in the context of protoplanetary disk dynamics. These are
related to planet-disk interactions in dusty disks
\citep{Benitez-Llambay2018}, dust filtration by giant planets
\citep{Weber2018} and dust accumulation in magnetized protoplanetary
disks when non-ideal MHD effects are taken into account
\citep{Krapp2018}.  The robustness of the implicit scheme presented in
this work and its versatility for adding an arbitrary number of
species in a systematic way, open new possibilities for studying
dust-dynamics self-consistently in protoplanetary disks.

These capabilities are critical in order to investigate a wide range
of phenomena in dusty protoplanetary disks, including dust growth,
dust and planetesimal dynamics and, ultimately how planets form.

\acknowledgements

We thank Fr\'ed\'eric Masset for inspiring
discussions that motivated this project.
We thank Philipp Weber for his valuable contribution based on an early
version of our numerical implementation. His input was key to define
the flowchart presented in this work.
We thank Richard Booth for his demonstration of the importance of the
order of the operator splitting when computing the implicit update,
discussed in Section \ref{subsec:implementation}.
We thank Andrew Youdin for suggesting us adding up densities when
computing cumulative density distributions in order to ease the
comparison between the fluid approach and Lagrangian particles.
This project has received funding from the European Union's Horizon
2020 research and innovation programme under grant agreement No 748544
(PBLL).  The research leading to these results has received funding
from the European Research Council under the European Union's Horizon
2020 research and innovation programme (grant agreement No 638596)
(LK). MEP gratefully acknowledges support from the Independent
Research Fund Denmark (DFF) via grant no.\,DFF
8021-00400B. Computations were performed on the \texttt{astro\_gpu}
and \texttt{astro\_long} partitions of the Steno cluster at the HPC
center of the University of Copenhagen.

\software{IPython \citep{Perez2007}, NumPy \citep{Walt2011}, Matplotlib \citep{Hunter2007}}


\begin{thebibliography}{}
	\expandafter\ifx\csname natexlab\endcsname\relax\def\natexlab#1{#1}\fi
	
	\bibitem[{{Bai} \& {Stone}(2010)}]{Bai2010}
	{Bai}, X.-N., \& {Stone}, J.~M. 2010, \apjs, 190, 297
	
	\bibitem[{{Balsara} {et~al.}(2009){Balsara}, {Tilley}, {Rettig}, \&
		{Brittain}}]{Balsara2009}
	{Balsara}, D.~S., {Tilley}, D.~A., {Rettig}, T., \& {Brittain}, S.~D. 2009,
	\mnras, 397, 24
	
	\bibitem[{{Barge} \& {Sommeria}(1995)}]{Barge1995}
	{Barge}, P., \& {Sommeria}, J. 1995, \aap, 295, L1
	
	\bibitem[{{Baruteau} \& {Zhu}(2016)}]{Baruteau2016}
	{Baruteau}, C., \& {Zhu}, Z. 2016, \mnras, 458, 3927
	
	\bibitem[{Benilov(1997)}]{Benilov1997}
	Benilov, M.~S. 1997, Physics of Plasmas, 4, 521
	
	\bibitem[{Ben{\'i}tez-Llambay \& Masset(2016)}]{Benitez-Llambay2016}
	Ben{\'i}tez-Llambay, P., \& Masset, F.~S. 2016, \apjs, 223, 11
	
	\bibitem[{{Ben{\'{\i}}tez-Llambay} \& {Pessah}(2018)}]{Benitez-Llambay2018}
	{Ben{\'{\i}}tez-Llambay}, P., \& {Pessah}, M.~E. 2018, \apjl, 855, L28
	
	\bibitem[{Ben{\'i}tez-Llambay {et~al.}(2016)Ben{\'i}tez-Llambay, Ramos,
		Beaug{\'e}, \& Masset}]{Benitez-Llambay2016a}
	Ben{\'i}tez-Llambay, P., Ramos, X.~S., Beaug{\'e}, C., \& Masset, F.~S. 2016,
	\apj, 826, 13
	
	\bibitem[{Berman \& Plemmons(1979)}]{Berman1979}
	Berman, A., \& Plemmons, R.~J. 1979, Nonnegative matrices in the mathematical
	sciences (Academic Press New York), xviii, 316 p.
	
	\bibitem[{{B{\'e}thune} {et~al.}(2016){B{\'e}thune}, {Lesur}, \&
		{Ferreira}}]{Bethune2016}
	{B{\'e}thune}, W., {Lesur}, G., \& {Ferreira}, J. 2016, \aap, 589, A87
	
	\bibitem[Booth et al.(2015)]{Booth2015} Booth, R.~A., Sijacki, D., \& Clarke, C.~J.\ 2015, \mnras, 452, 3932 
	
	\bibitem[{{Braginskii}(1965)}]{Braginskii1965}
	{Braginskii}, S.~I. 1965, Reviews of Plasma Physics, 1, 205
	
	\bibitem[{{Brandenburg} \& {Dobler}(2002)}]{Brandenburg2002}
	{Brandenburg}, A., \& {Dobler}, W. 2002, Computer Physics Communications, 147,
	471
	
	\bibitem[{{Chen} \& {Lin}(2018)}]{Chen2018}
	{Chen}, J.-W., \& {Lin}, M.-K. 2018, \mnras, 478, 2737
	
	\bibitem[{Chrenko {et~al.}(2017)Chrenko, Bro{\v z}, \&
		Lambrechts}]{Chrenko2017}
	Chrenko, O., Bro{\v z}, M., \& Lambrechts, M. 2017, \aap, 606, A114
	
	\bibitem[{de~Val-Borro {et~al.}(2006)de~Val-Borro, Edgar, Artymowicz,
		Ciecielag, Cresswell, D'Angelo, Delgado-Donate, Dirksen, Fromang,
		Gawryszczak, Klahr, Kley, Lyra, Masset, Mellema, Nelson, Paardekooper,
		Peplinski, Pierens, Plewa, Rice, Sch{\"{a}}fer, \& Speith}]{Val-Borro2006}
	de~Val-Borro, M., Edgar, R., Artymowicz, P., {et~al.} 2006, \mnras, 370, 529
	
	\bibitem[{Dipierro \& Laibe(2017)}]{Dipierro2017}
	Dipierro, G., \& Laibe, G. 2017, \mnras, 469, 1932
	
	\bibitem[{{Dipierro} {et~al.}(2018){Dipierro}, {Laibe}, {Alexander}, \&
		{Hutchison}}]{Dipierro2018}
	{Dipierro}, G., {Laibe}, G., {Alexander}, R., \& {Hutchison}, M. 2018, \mnras,
	479, 4187
	
	\bibitem[{Dr{\c a}zkowska {et~al.}(2010)Dr{\c a}zkowska, Hanasz, \&
		Kowalik}]{Drcakowska2010}
	Dr{\c a}zkowska, J., Hanasz, M., \& Kowalik, K. 2010, Particle module of
	Piernik MHD code, Astrophysics Source Code Library, , , ascl:1010.005
	
	\bibitem[{Fromang \& Papaloizou(2006)}]{Fromang2006b}
	Fromang, S., \& Papaloizou, J. 2006, \aap, 452, 751
	
	\bibitem[{Fung {et~al.}(2014)Fung, Shi, \& Chiang}]{Fung2014}
	Fung, J., Shi, J.-M., \& Chiang, E. 2014, \apj, 782, 88
	
	\bibitem[{{Hanasz} {et~al.}(2010){Hanasz}, {Kowalik}, {W{\'o}lta{\'n}ski},
		{Paw{\l}aszek}, \& {Kornet}}]{Hanasz2010}
	{Hanasz}, M., {Kowalik}, K., {W{\'o}lta{\'n}ski}, D., {Paw{\l}aszek}, R., \&
	{Kornet}, K. 2010, in EAS Publications Series, Vol.~42, EAS Publications
	Series, ed. K.~{Go{\.z}dziewski}, A.~{Niedzielski}, \& J.~{Schneider},
	281--285
	
	\bibitem[{{Hawley} {et~al.}(1984){Hawley}, {Smarr}, \& {Wilson}}]{Hawley1984}
	{Hawley}, J.~F., {Smarr}, L.~L., \& {Wilson}, J.~R. 1984, \apjs, 55, 211
	
	\bibitem[{Hunter(2007)}]{Hunter2007}
	Hunter, J.~D. 2007, Computing In Science \& Engineering, 9, 90
	
	\bibitem[{Hutchison {et~al.}(2018)Hutchison, Price, \& Laibe}]{Hutchison2018}
	Hutchison, M., Price, D.~J., \& Laibe, G. 2018, Monthly Notices of the Royal
	Astronomical Society, 476, 2186
	
	\bibitem[{Johansen {et~al.}(2004)Johansen, Andersen, \&
		Brandenburg}]{Johansen2004}
	Johansen, A., Andersen, A.~C., \& Brandenburg, A. 2004, \aap, 417, 361
	
	\bibitem[{Johansen \& Youdin(2007)}]{JY07}
	Johansen, A., \& Youdin, A. 2007, The Astrophysical Journal, 662, 627
	
	\bibitem[{Kowalik {et~al.}(2013)Kowalik, Hanasz, W{\'o}lta{\'n}ski, \&
		Gawryszczak}]{Kowalik2013a}
	Kowalik, K., Hanasz, M., W{\'o}lta{\'n}ski, D., \& Gawryszczak, A. 2013,
	\mnras, 434, 1460
	
	\bibitem[{{Krapp} {et~al.}(2018){Krapp}, {Gressel}, {Ben{\'{\i}}tez-Llambay},
		{Downes}, {Mohandas}, \& {Pessah}}]{Krapp2018}
	{Krapp}, L., {Gressel}, O., {Ben{\'{\i}}tez-Llambay}, P., {et~al.} 2018, \apj,
	865, 105
	
	\bibitem[{{Laibe} \& {Price}(2011)}]{Laibe2011}
	{Laibe}, G., \& {Price}, D.~J. 2011, \mnras, 418, 1491
	
	\bibitem[{{Laibe} \& {Price}(2012)}]{Laibe2012a}
	---. 2012, \mnras, 420, 2345
	
	\bibitem[{Laibe \& Price(2014)}]{Laibe2014}
	Laibe, G., \& Price, D.~J. 2014, Monthly Notices of the Royal Astronomical
	Society, 444, 1940
	
	\bibitem[{{Lehmann} \& {Wardle}(2018)}]{Lehmann2016}
	{Lehmann}, A., \& {Wardle}, M. 2018, \mnras, 476, 3185
	
	\bibitem[{Lyra \& Lin(2013)}]{Lyra2013}
	Lyra, W., \& Lin, M.-K. 2013, \apj, 775, 17
	
	\bibitem[{Miniati(2010)}]{Miniati2010}
	Miniati, F. 2010, Journal of Computational Physics, 229, 3916
	
	\bibitem[{{Nakagawa} {et~al.}(1986){Nakagawa}, {Sekiya}, \&
		{Hayashi}}]{Nakagawa1986}
	{Nakagawa}, Y., {Sekiya}, M., \& {Hayashi}, C. 1986, \icarus, 67, 375
	
	\bibitem[{Paardekooper \& Mellema(2006)}]{Paardekooper2006a}
	Paardekooper, S.-J., \& Mellema, G. 2006, aap, 453, 1129
	
	\bibitem[{P\'erez \& Granger(2007)}]{Perez2007}
	P\'erez, F., \& Granger, B.~E. 2007, Computing in Science and Engineering, 9,
	21
	
	\bibitem[{{Porth} {et~al.}(2014){Porth}, {Xia}, {Hendrix}, {Moschou}, \&
		{Keppens}}]{MPIAMRVAC}
	{Porth}, O., {Xia}, C., {Hendrix}, T., {Moschou}, S.~P., \& {Keppens}, R. 2014,
	\apjs, 214, 4
	
	\bibitem[{Press {et~al.}(2007)Press, Teukolsky, Vetterling, \&
		Flannery}]{Press2007}
	Press, W.~H., Teukolsky, S.~A., Vetterling, W.~T., \& Flannery, B.~P. 2007,
	Numerical Recipes 3\textsuperscript{rd} Edition: The Art of Scientific
	Computing, 3rd edn. (New York, NY, USA: Cambridge University Press)
	
	\bibitem[{{Price} {et~al.}(2018){Price}, {Wurster}, {Tricco}, {Nixon},
		{Toupin}, {Pettitt}, {Chan}, {Mentiplay}, {Laibe}, {Glover}, {Dobbs},
		{Nealon}, {Liptai}, {Worpel}, {Bonnerot}, {Dipierro}, {Ballabio}, {Ragusa},
		{Federrath}, {Iaconi}, {Reichardt}, {Forgan}, {Hutchison}, {Constantino},
		{Ayliffe}, {Hirsh}, \& {Lodato}}]{Price2018}
	{Price}, D.~J., {Wurster}, J., {Tricco}, T.~S., {et~al.} 2018, \pasa, 35, e031
	
	\bibitem[{{Ragusa} {et~al.}(2017){Ragusa}, {Dipierro}, {Lodato}, {Laibe}, \&
		{Price}}]{Ragusa2017}
	{Ragusa}, E., {Dipierro}, G., {Lodato}, G., {Laibe}, G., \& {Price}, D.~J.
	2017, \mnras, 464, 1449
	
	\bibitem[{{Riols} \& {Lesur}(2018)}]{Riols2018}
	{Riols}, A., \& {Lesur}, G. 2018, \aap, 617, A117
	
	\bibitem[{Safronov(1972)}]{Safronov1972}
	Safronov, V. 1972, Evolution of the protoplanetary cloud and formation of the
	earth and planets.
	
	\bibitem[{Stone {et~al.}(2008)Stone, Gardiner, Teuben, Hawley, \&
		Simon}]{Stone2008}
	Stone, J., Gardiner, T., Teuben, P., Hawley, J., \& Simon, J. 2008, \apjs, 178,
	137
	
	\bibitem[{Stone \& Norman(1992)}]{Stone1992}
	Stone, J., \& Norman, M. 1992, \apjs, 80, 753
	
	\bibitem[{{Stone}(1997)}]{Stone1997}
	{Stone}, J.~M. 1997, \apj, 487, 271
	
	\bibitem[{{Stoyanovskaya} {et~al.}(2018){Stoyanovskaya}, {Glushko},
		{Snytnikov}, \& {Snytnikov}}]{Stoyanovskaya2018b}
	{Stoyanovskaya}, O.~P., {Glushko}, T.~A., {Snytnikov}, N.~V., \& {Snytnikov},
	V.~N. 2018, Astronomy and Computing, 25, 25
	
	\bibitem[{Stoyanovskaya {et~al.}(2018)Stoyanovskaya, Vorobyov, \&
		Snytnikov}]{Stoyanovskaya2018}
	Stoyanovskaya, O.~P., Vorobyov, E.~I., \& Snytnikov, V.~N. 2018, Astronomy
	Reports, 62, 455
	
	\bibitem[{Vorobyov {et~al.}(2018)Vorobyov, Akimkin, Stoyanovskaya,
		Pavlyuchenkov, \& Liu}]{Vorobyov2018}
	Vorobyov, E.~I., Akimkin, V., Stoyanovskaya, O., Pavlyuchenkov, Y., \& Liu,
	H.~B. 2018, \aap, 614, A98
	
	\bibitem[{Walt {et~al.}(2011)Walt, Colbert, \& Varoquaux}]{Walt2011}
	Walt, S. v.~d., Colbert, S.~C., \& Varoquaux, G. 2011, Computing in Science and
	Engg., 13, 22
	
	\bibitem[{Weber {et~al.}(2018)Weber, Ben{\'{i}}tez-Llambay, Gressel, Krapp, \&
		Pessah}]{Weber2018}
	Weber, P., Ben{\'{i}}tez-Llambay, P., Gressel, O., Krapp, L., \& Pessah, M.
	2018, \apj, 854, 153
	
	\bibitem[{Weidenschilling(1977)}]{Weidenschilling1977}
	Weidenschilling, S.~J. 1977, \mnras, 180, 57
	
	\bibitem[{{Whipple}(1972)}]{Whipple1972}
	{Whipple}, F.~L. 1972, in From Plasma to Planet, ed. A.~{Elvius}, 211
	
	\bibitem[{{Yang} \& {Johansen}(2016)}]{Yang2016}
	{Yang}, C.-C., \& {Johansen}, A. 2016, \apjs, 224, 39
	
	\bibitem[{Youdin \& Johansen(2007)}]{Youdin2007}
	Youdin, A., \& Johansen, A. 2007, \apj, 662, 613
	
	\bibitem[{Youdin \& Goodman(2005)}]{Youdin2005}
	Youdin, A.~N., \& Goodman, J. 2005, \apj, 620, 459
	
	\bibitem[{{Zhu} \& {Stone}(2014)}]{Zhu2014}
	{Zhu}, Z., \& {Stone}, J.~M. 2014, \apj, 795, 53
	
\end{thebibliography}

\appendix

\section{Source terms in the continuity equation}

\label{ap:source_term}

Important physical processes can be modeled by means of source terms
in the continuity equation. Some examples are dust
fragmentation/growth, dust diffusion, and chemical reactions, among others
\citep[see e.g.][for an example on dust diffusion]{Weber2018}.
The continuity equation with a source term (possibly depending on
other species) is
\begin{equation}
\frac{\partial \rho_{i}}{\partial t} +  \nabla \cdot \left( \rho_{i} {\bf v}_i\right)  = S_i\,.
\end{equation}

It is important to note that the source terms in the continuity
equation must appear as the source terms $S_i {\bf v}_i$ in the
conservative form of the momentum equation for each species. To account for the source
terms in both the continuity and momenta equations, the transport step
can be split into two sub-steps applying the operator splitting
technique.

\section{Stability and convergence of the implicit scheme} 
\label{ap:NS}

In Section \ref{sec:numericalmethod} we addressed the stability
and convergence of the implicit scheme based on the fact that ${\bf T}^{-1}$ exist, 
it is diagonalizable, strictly positive, and right
stochastic.
In this appendix we demonstrate these properties.

\subsection{${\bf T}$ is nonsingular}
\label{ap:eigenvalues}

To show that ${\bf T}$ is nonsingular, it is enough to prove that zero
is not an eigenvalue of it. We first note that, if $\lambda_T$ is an eigenvalue of ${\bf T} = {\bf
  I}+ \Delta t {\bf M}$, then $\lambda_M = \left(\lambda_T -
1\right)/\Delta t$ is an eigenvalue of ${\bf M}$.  Upper and lower
bounds for $\lambda_M$ can be found by means of the Gershgorin circle
theorem which, from Eq.\,\eqref{eq:M}, implies
\begin{equation}
  0\leq \lambda_{M} \leq 2 \max_{k=1,\dots,N}\left( \sum_{j\neq k}^N\alpha_{kj} \right)\,.
\end{equation}
Hence, all the eigenvalues of ${\bf M}$ are real and non-negative. Since
\begin{equation}
  \sum_{k=1}^N {\bf M}_{ik} = 0 \quad \, \textrm{for every }  i=1,\dots,N\,.
\end{equation} 
$\lambda_{M} = 0$ is an eigenvalue and ${\bf M}$ is singular. We then
conclude that $\lambda_T \geq 1$ provided $\Delta t>0$.

\subsection{${\bf T}^{-1}$ is diagonalizable}

\label{ap:diagonal}

Since ${\bf M}$ and ${\bf T}$ commute they are simultaneously
diagonalizable. Therefore, to prove that ${\bf T}^{-1}$ is
diagonalizable, it is enough to show that ${\bf M}$ is similar to a
(real) symmetric matrix, i.e., ${\bf S} = {\bf D}^{-1} {\bf M} {\bf
  D}$ with ${\bf S} = {\bf S}^{\mathsf{T}}$. We demonstrate this as
follows. Defining the diagonal matrix, ${\bf R}$, with elements
$R_{ij} = \rho_i/\rho_0 \delta_{ij}$, for any arbitrary $\rho_0$ and
$i,j = 1, \ldots N$, Eq.\,\eqref{eq:symmetry} implies $R_{ik} M_{kj} =
R_{jk} M_{ki}$, i.e., ${\bf RM} = ({\bf RM})^{\mathsf{T}}$ is
symmetric. Because ${\bf R}$ is diagonal it follows that ${\bf RM} =
{\bf M}^{\mathsf{T}} {\bf R}$. Multiplying this last equality, to both
left and right, by the matrix ${\bf R}^{-1/2}$, we obtain ${\bf
  R}^{1/2} {\bf M} {\bf R}^{-1/2} = {\bf R}^{-1/2} {\bf
  M}^{\mathsf{T}} {\bf R}^{1/2} = ({\bf R}^{1/2} {\bf M} {\bf
  R}^{-1/2})^{\mathsf{T}}$. This demonstrates that the matrix ${\bf S}
= {\bf D} ^{-1} {\bf M} {\bf D} $ with ${\bf D} = {\bf R}^{-1/2}$ is
symmetric.

\subsection{${\bf T}^{-1}$ is right stochastic}

\label{ap:stochastic}

A right stochastic matrix is defined as a matrix whose entries are
non-negative and with each row summing to one.

From Eq.\,\eqref{eq:M}, it is clear that
\begin{align}
  \sum_{j=1}^{N} {\bf T}_{ij} = 1 + \Delta t \sum_{j=1}^{N} {\bf M}_{ij} = 1\,,
\end{align}
and because $\sum_j \sum_k T_{ik}T^{-1}_{kj} = \sum_j \delta_{ij}$, $\sum_j {\bf T}_{ij}^{-1}=1$. Furthermore,
\begin{equation}
  {\bf T}_{ij} < 0\,, \quad \textrm{for } i\neq j\,,
\end{equation}
so ${\bf T}$ belongs to the group of matrices $\mathbb{Z}^{n\times
  n}$. In addition, since the eigenvalues of ${\bf T}$ are real and
positive (see Appendix \ref{ap:eigenvalues}), ${\bf T}$ is a
non-singular $\mathbb{M}$-matrix \citep[see Theorem 2.3, chapter
  6][]{Berman1979} and ${\bf T}$ has a non-negative inverse.

\subsection{${\bf T}^{-1}$ is strictly positive}
\label{ap:positive}

The inverse, ${\bf T}^{-1}$, of an irreducible non-singular
$\mathbb{M}$-matrix is strictly positive \citep[see Theorem 2.7,
  chapter 6][]{Berman1979}, i.e., ${\bf T}^{-1}_{i,j} > 0$ for all
$i,j \in 1 \dots N$.
Since ${\bf T}$ is a nonsingular $\mathbb{M}$-matrix (see Appendix
\ref{ap:stochastic}), it only remains to be demonstrated that ${\bf
  T}$ is irreducible.

Theorem 2.7 (chapter 2) from \cite{Berman1979}, states that a matrix
${\bf T}$ is irreducible if and only if its direct graph, $G({\bf
  T})$, is strongly connected. $G({\bf T})$ consists of $n$ vertices
$P_1, \dotsc, P_n$, where an edge leads from $P_i$ to $P_j$ if and
only if $T_{ij}\neq0$. Furthermore, it is strongly connected if for
any ordered pair $(P_i,P_j)$ of vertices, there exist a path that
leads from $P_i$ to $P_j$. Clearly, if all the species collide with
each other, then $T_{ij} \neq 0$ for all $i,j$ and $G({\bf T})$ is
strongly connected. This statement holds, even when neglecting direct
collisions between species $i$ and $j$, provided that each species
collides with another (proxy) species $k$. This is because the path
from $P_i$ to $P_j$ is defined through $P_k$.

\section{Non-linear drag force}
\label{ap:NL}

The collision rate might depend on the relative velocity between
species, $\Delta v$, and Eqs.\,\eqref{eq:new_source} become non-linear
in the velocities. Since the implicit scheme described by
Eq.\,\eqref{eq:implicit_form} assumes a linear system, in principle,
it cannot be used to find the solution in such a non-linear regime. In
this appendix, we present simple tests of an approximation that allows
us to circumvent this issue, which consists in assuming the system to
be linear in $\Delta v^{n+1}$, and the collision rate, $\alpha^n$, to
be dependent on $\Delta v^{n}$. This approximation makes it possible
for all the properties of the implicit scheme hold even in the
non-linear drag regime. By mean of a simple example, we show that this
method is good enough to recover the solution of
Eqs.\,\eqref{eq:new_source} in the non-linear regime. For this test we
adopt the three different collision rates presented in the
Table\,\ref{tab:damping_nl}.
\begin{table}[h!]
	\begin{center}
		\caption{Collision rate $\alpha$, for the different non-linear
			drag force laws showed in Fig\,\ref{fig:damp_nl}. The coefficients
			$\gamma$, $\gamma_q$, $\gamma_p$, $\gamma_m$ and $b_m$ are all
			fixed to one for the purpose of this test.}
		\label{tab:damping_nl}
		\begin{tabular}{cccc}
			\decimals
			\hline\hline
			Linear   & Quadratic & Power-law & Mixed \\
			\hline
			$\gamma$ & $\gamma_q |\Delta v|$ & $\gamma_p |\Delta v|^{1/p}$ & $\gamma_m \sqrt{1+b_m|\Delta v|^2}$ \\
			\hline
			\hline
		\end{tabular}
	\end{center}
\end{table}

We consider two species with initial densities and velocities, $\rho_1
= \rho_2 = 1$ and $v_1 = 20$, $v_2 = 10$, respectively, and obtain
numerical solutions following Section \ref{sec:damping}. For the
power-law regime, we set the index $p=2$.

\begin{figure}[h!]
	\centering
	\includegraphics[]{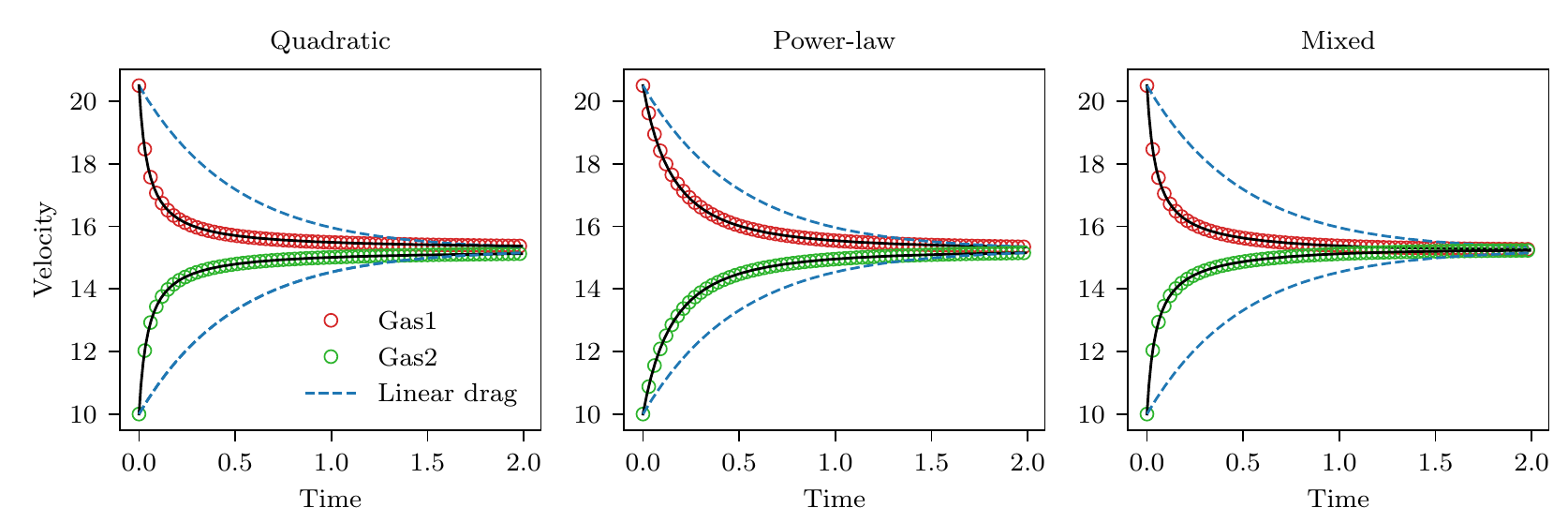}
	\caption{Analytical (solid lines) and numerical (open circles)
		solutions for a quadratic (left panel), power law with $p=2$
		(center panel) and mixed (right panel) drag forces. The dashed
		line is the solution for a linear drag force.}\label{fig:damp_nl}
\end{figure}

In each panel of Fig.\,\ref{fig:damp_nl} we show the analytical and
numerical solutions, with solid lines and open circles,
respectively. The dashed lines correspond to the solution obtained for
a linear drag force. In all the non-linear drag regimes, the agreement
between the analytical and numerical solutions is excellent.

\newpage

\section{Eigenvalues for the sound wave test problem} \label{ap:dispersion-soundwave}

To solve the dispersion relation \eqref{eq:dispersion}, we write it as the polynomial
equation
\begin{equation}
P(\omega) = \omega^{N+1} + a_{N}\omega^{N} + \dotsc + a_{0} = 0\,.
\label{eq:polynomial}
\end{equation}
The $N+1$ roots of \eqref{eq:polynomial} are the eigenvalues of the
problem. The polynomial $P$ has at least $N-1$ real positive
roots. This is proved by first noticing that $F$ is positive for
$\omega < \min\left(t_{{\rm s}m}^{-1}\right)$. Since $t_{{\rm
		s}m}>0$, no real root exists for $\omega < 0$. In addition, the
one-sided limits of $F$ satisfy
\begin{align}
\lim_{\omega\to t_{{\rm s}m}^{-1 \pm}} F(\omega,\omega_{\rm s}) = \pm \infty\,,
\end{align}
i.e., the function changes sign at each side of the singular points,
from which we conclude that there is at least one real positive root
between two adjacent singular points, giving $N-2$ positive
roots. Furthermore, since
\begin{align}
\lim_{\omega\to \infty} F(\omega,\omega_{\rm s}) = \infty\,,
\end{align}
there is at least one more positive root beyond the last singular
point. All of these roots correspond to pure damping solutions.

An upper bound for the largest real root can found. Defining $t_{{\rm
    s}N-1}^{-1}$ as the largest singular point, for $\omega \gg
t_{{\rm s}N-1}^{-1}$, $f>0$ if
\begin{equation}
1+\sum_{m=1}^{N-1}\frac{\epsilon_m}{1-\omega t_{{\rm s}m}} \simeq 1-\frac{1}{\omega} \sum_{m=1}^{N-2}\frac{\epsilon_m}{t_{{\rm s}m}} + \frac{\epsilon_{N-1}}{1-\omega t_{{\rm s}N-1}} > 0\,.
\end{equation}
This expression is equivalent to the quadratic inequality
\begin{equation}
-t_{{\rm s}N-1}\omega^2 + \left[1 + \epsilon_{N-1} + t_{{\rm s}N-1}\sum_{m=1}^{N-2}\frac{\epsilon_m}{t_{{\rm s}m}} \right]\omega -  \sum_{m=1}^{N-2}\frac{\epsilon_m}{t_{{\rm s}m}} > 0\,,\label{eq:quadratic}
\end{equation}
from which we find that the largest positive root of $P$ is smaller
than the largest root of \eqref{eq:quadratic}.

Having found the $N-1$ roots of $P$, $\{\omega_1, \dots, \omega_{N-1} \}$, we can use Vieta's formulae
to find the remaining roots $\omega_{N}$ and $\omega_{N+1}$
\begin{align}
\omega_{N}\omega_{N+1} &= \left(-1\right)^{N+1}a_0 \prod_{j=1}^{N-1} \omega_j^{-1} \label{eq:lambdaN1}\,, \\
\omega_{N} + \omega_{N+1} &= -a_N - \sum_{j=1}^{N-1} \omega_j\,. \label{eq:lambdaN2}
\end{align}
where $a_0, a_{N}$ are
\begin{align}
a_0 &= \left(-1\right)^{N-1} \omega_{\rm s}^2 \prod_{j=1}^{N-1} t_{{\rm s}j}^{-1}\,, \\
a_{N} &= -\sum_{j=1}^{N-1} t_{{\rm s}j}^{-1} \left( 1 + \epsilon_j\right)\,.
\end{align}
Eqs.\,\eqref{eq:lambdaN1}-\eqref{eq:lambdaN2} can be written as a
second order polynomial equation, from which we obtain the final two
roots, which are, in general, complex, and are the most interesting
ones.

\newpage

\section{Equations for the streaming instability} \label{ap:SI}

We linearize
Eqs.\,\eqref{eq:streaming_gas_continuity}-\eqref{eq:streaming_dust_momenta}
around the background solution
\eqref{eq:equi_gas}-\eqref{eq:equi_dust}. Without loss of generality,
we assume perturbations of the form $\delta f = \delta \hat{f}
e^{i(k_x x + k_z z) - \omega t}$. Defining the dimensionless densities
$\delta \tilde{\rho} = \delta \hat{\rho}/\rho^0_{{\rm g}}$,
$\epsilon^0_k = \rho_{k}^0/\rho_{\rm{g}}^0$, velocities $\tilde{v} =
\hat{v}/(h_0^2v_{{\rm K}0})$, wavenumber $K = k h_0^2 v_{{\rm
    K}0}/\Omega_0$, eigenvalue $\tilde \omega = \omega/\Omega_0$, and
relative velocities $\Delta^0_{kx} = {\tilde{v}^0_{{\rm g}x} -
  \tilde{v}^0_{kx}}$, $\Delta^0_{ky} = {\tilde{v}^0_{{\rm g}y} -
  \tilde{v}^0_{ky}}$, the equations describing the linear evolution of
the system are

\begin{align}
iK_x\tilde{v}_{{\rm g}x}^{0} \delta \tilde{\rho}_{\rm g} + iK_x \delta\tilde{v}_{{\rm g}x} + iK_z \delta \tilde{v}_{{\rm g}z} &= \tilde{\omega} \delta \tilde{\rho}_{\rm g}\\
\left(iK_xh_0^{-2} - \sum_{k=1}^{N}\frac{\epsilon^0_k {\Delta}^0_{kx}}{T_{{\rm s}k}}\right)\delta \tilde{\rho}_{\rm g} +
\left(iK_x\tilde{v}_{{\rm g}x}^{0} + \sum_{k=1}^{N} \frac{\epsilon^0_k}{T_{{\rm s}k}}\right)\delta \tilde{v}_{{\rm g}x}  -
2\delta \tilde{v}_{{\rm g} y} +
\sum_{k=1}^{N}\frac{{\Delta}^0_{kx}}{T_{{\rm s}k}} \delta \tilde{\rho}_k - \sum_{k=1}^{N}\frac{\epsilon^0_k}{T_{{\rm s}k}} \delta \tilde{v}_{kx}
&= \tilde{\omega} \delta \tilde{v}_{{\rm g}x} \\
-\left( \sum_{k=1}^{N}\frac{\epsilon^0_k {\Delta}^0_{ky}}{T_{{\rm s}k}}\right) \delta \tilde{\rho}_{\rm g}+
\left(2-q\right) \delta \tilde{v}_{{\rm g}x} + 
\left(iK_x\tilde{v}_{{\rm g}x}^{0} + \sum_{k=1}^{N} \frac{\epsilon^0_k}{T_{{\rm s}k}}\right)\delta \tilde{v}_{{\rm g}y}  +
\sum_{k=1}^{N}\frac{{\Delta}^0_{ky}}{T_{{\rm s}k}} \delta \tilde{\rho}_k - \sum_{k=1}^{N}\frac{\epsilon^0_k}{T_{{\rm s}k}} \delta \tilde{v}_{ky} 
&= \tilde{\omega} \delta \tilde{v}_{{\rm g}y} \\
iK_zh_0^{-2}\delta \tilde{\rho}_{\rm g} + 
\left(iK_x\tilde{v}_{{\rm g}x}^{0} + \sum_{k=1}^{N} \frac{\epsilon^0_k}{T_{{\rm s}k}}\right)\delta \tilde{v}_{{\rm g}z} -
\sum_{k=1}^{N}\frac{\epsilon^0_k}{T_{{\rm s}k}} \delta \tilde{v}_{kz}
&= \tilde{\omega} \delta \tilde{v}_{{\rm g}z} \\
iK_x\tilde{v}_{jx}^{0} \delta \tilde{\rho}_{j} + iK_x\epsilon^0_{j}\delta\tilde{v}_{jx} + iK_z\epsilon^0_{j} \delta\tilde{v}_{jz} &= \tilde{\omega} \delta \tilde{\rho}_{j} \\
-\frac{1}{T_{{\rm s}j}} \delta \tilde{v}_{{\rm g} x}+
\left(iK_x\tilde{v}_{jx}^{0} + \frac{1}{T_{{\rm s}j}}\right)\delta \tilde{v}_{jx} -
2\delta \tilde{v}_{j y}
&= \tilde{\omega} \delta \tilde{v}_{jx} \\
- \frac{1}{T_{{\rm s}j}} \delta \tilde{v}_{{\rm g} y} +
\left(2-q\right) \delta \tilde{v}_{j x} + 
\left(iK_x\tilde{v}_{jx}^{0} + \frac{1}{T_{{\rm s}j}}\right) \delta \tilde{v}_{jy}
&= \tilde{\omega} \delta \tilde{v}_{jy} \\
-\frac{1}{T_{{\rm s}j}} \delta \tilde{v}_{{\rm g} z} +
\left(iK_x\tilde{v}_{jx}^{0} + \frac{1}{T_{{\rm s}j}}\right) \delta \tilde{v}_{jz}
&= \tilde{\omega} \delta \tilde{v}_{jz}
\end{align}

\end{document}